\documentclass[11pt]{iopart}

\usepackage{amssymb}
\usepackage{ifpdf}
 \usepackage{enumerate}
\usepackage{graphicx}
\usepackage{hyperref}

\hypersetup{
    pdfnewwindow=true,      
    colorlinks=true,       
    linkcolor=blue,          
    citecolor=blue,        
    filecolor=blue,      
    urlcolor=blue           
}

\providecommand{\aj}[0]{Astron. J.}
\providecommand{\apj}[0]{Astrophys. J.}
\providecommand{\apjl}[0]{Astrophys. J. Lett.}

\providecommand{\aap}[0]{Astron. Astrophys. }

\providecommand{\araa}[0]{Ann.\ Rev. Astron. Astroph. }
\providecommand{\physrep}[0]{Phys. Rep. }
\providecommand{\mnras}[0]{Mon. Not. Roy. Astron. Soc. }

\providecommand{\nat}[0]{Nature}
\providecommand{\prl}[0]{Phys. Rev. Lett.}
\providecommand{\prd}{Phys. Rev. D.}

\providecommand{\nar}[0]{New Astron. Rev.}
\providecommand{\physrep}[0]{Phys. Rep.}
\providecommand{\ssr}[0]{Space Sci. Rev.}

\def\imbh#1{intermediate mass black hole#1(IMBH#1)\gdef\imbh{IMBH}}
\def\smbh#1{supermassive black hole#1(SMBH#1)\gdef\smbh{SMBH}}
\def\bbh#1{binary black hole#1 (BBH#1)\gdef\bbh{BBH}}
\def\bh#1{black hole#1 (BH#1)\gdef\bh{BH}}
\def\ns#1{neutron star#1 (NS#1)\gdef\ns{NS}}
\def\gw#1{gravitational wave#1 (GW#1)\gdef\gw{GW}}
\def\pnw#1{post-Newtonian#1 (PN#1)\gdef\pnw{PN}}
\def\eos#1{equation of state#1 (EOS#1)\gdef\eos{EOS}}

\begin{document}

\title{How Gravitational-wave Observations Can Shape the Gamma-ray Burst Paradigm}
\date{\today}

\author{I.~Bartos$^{1}$,
        P.~Brady$^{2}$,
        S.~M\'arka$^{1}$}
\address{$^{1}$Department of Physics \& Columbia Astrophysics Laboratory, Columbia University, New York, NY 10027, USA}
\address{$^{2}$Center for Gravitation and Cosmology, University of Wisconsin-Milwaukee, Milwaukee, WI 53211, USA}

\begin{abstract}
By reaching through shrouding blastwaves, efficiently discovering off-axis events, and probing the central engine at work, gravitational wave (GW) observations will soon revolutionize the study of gamma-ray bursts. Already, analyses of GW data targeting gamma-ray bursts have helped constrain the central engines of selected events. Advanced GW detectors with significantly improved sensitivities are under construction. After outlining the GW emission mechanisms from gamma-ray burst progenitors (binary coalescences, stellar core collapses, magnetars, and others) that may be detectable with advanced detectors, we review how GWs will improve our understanding of gamma-ray burst central engines, their astrophysical formation channels, and the prospects and methods for different search strategies. We place special emphasis on multimessenger searches. To achieve the most scientific benefit, GW, electromagnetic, and neutrino observations should be combined to provide greater discriminating power and science reach.
\end{abstract}



\maketitle
\tableofcontents

\section{Introduction}

Gamma-ray bursts (GRBs) are the brightest electromagnetic explosions in the Universe (e.g., \cite{Meszaros2012}). For the short time of their activity, they outshine all other sources in the sky in gamma rays. These energetic explosions originate from cataclysmic cosmic events, whose ``inner engine" that drives the observed emission is confined to volumes mere tens of kilometers across, as indicated by the duration and variability of gamma-ray emission (e.g., \cite{1999PhR...314..575P}). This inner engine is hidden from direct electromagnetic observations, even though the variability of the emission is directly related to the engine's activity. Gravitational waves (GWs), on the other hand, carry information directly from the inner engine to the observer.

With the near completion of advanced GW detectors \cite{2010CQGra..27h4006H,2011CQGra..28k4002A,2006CQGra..23S.207W,2011IJMPD..20.1755K}, we will soon be able to directly probe the dynamical processes leading to the creation of GRBs, which are less (or not) accessible via other messengers.
The Advanced LIGO \cite{2010CQGra..27h4006H} and Virgo \cite{AdV} detectors are planned to begin observation in 2015, albeit initially below design sensitivity, gradually reaching their design sensitivity around the end of the decade \cite{LVCcommissioning}. Upon reaching their design sensitivity, the Advanced LIGO and Virgo detectors are expected to be about 10 times more sensitive (at their most sensitive frequency band around $\sim150\,$Hz) than the initial LIGO/Virgo detectors. Further, advanced detectors will have a wider sensitive frequency band, reaching down to $\sim10\,$Hz. This will improve their sensitivity to wide-band sources such as compact binaries. The global GW detector network will further increase its reach in both distance and source direction reconstruction by the addition of further detectors \cite{2012arXiv1210.6362N}, such as the Japanese KAGRA (formerly LCGT; \cite{2011IJMPD..20.1755K}) detector that is under construction, as well as the planned third LIGO observatory in India \cite{LVCcommissioning}.

There are a number of mechanisms conjectured to result in a GRB.
Potential mechanisms include the formation of a central object surrounded by an accretion disk from the merger of binary neutron stars \cite{1984PAZh...10..422B,1986ApJ...308L..43P,1989Natur.340..126E} or a neutron star and a low-mass black hole \cite{2007PhR...442..166N}, the collapse of the core of a massive star \cite{Wo:93,2006ARAA..44..507W,hjorth:11}, the global reconfiguration of the magnetic fields in magnetized neutron stars \cite{1992ApJ...392L...9D,1992AcA....42..145P,2007PhR...442..166N}, the accretion-induced collapse of white dwarfs following accretion from a non-degenerate companion \cite{2008MNRAS.385.1455M,2009arXiv0908.1127M} or following the merger of a white dwarf binary \cite{2004ApJ...601L.167M,2009arXiv0908.1127M}, or even cosmic strings \cite{1987ApJ...316L..49B}. Identifying the mechanisms behind the creation of individual GRBs will be greatly aided by GW detections.

This review intends to survey the prospects of GW measurements in understanding the central engines of GRBs. While our theoretical view of GW emission mechanisms connected to GRBs is rapidly evolving, the main directions, the opportunities and limitations of observational GW astrophysics are becoming clearer.
With this review we aim to provide guidance to (i) astronomers interested in utilizing the capabilities of GW measurements to complement electromagnetic or neutrino observations in understanding GRBs, and (ii) GW scientists interested in the open questions in GRB astrophysics that can be addressed through GW measurements.

Compact binary mergers, the likely progenitors of most short GRBs (e.g., \cite{2007PhR...442..166N,2011NewAR..55....1B}, although see \cite{2008MNRAS.385.1455M} for alternative models), are one of the primary targets of GW searches (e.g., \cite{2009PhRvD..79l2001A}). Advanced GW detectors will be able to, among others, measure the properties of the binary \cite{1994PhRvD..49.2658C}, directly probe the dynamics of the central engine, and correlate the binary's masses and orientation with prompt and delayed electromagnetic counterparts.  These observations will constrain the central engine mechanism, provide standard sirens for measuring cosmological parameters \cite{1986Natur.323..310S,1993ApJ...411L...5C,1996PhRvD..53.2878F}, and constrain the nature of nuclear matter (see Section \ref{section:cbcphysics}).

GW emission from isolated central engines, such as stellar core collapses, accretion induced collapses or magnetars, is typically at higher frequencies and at lower amplitudes than for compact binaries. GW emission depends on the development of varying quadrupole moment, which is highly source and model dependent. The detection of GWs from isolated objects would help us understand the internal processes in massive stellar collapses, including the development of differentially rotating protoneutron stars, the nature of nuclear matter, or the creation of massive accretion disks (Sections \ref{section:collapsarphysics}, \ref{section:magnetarphysics} \& \ref{section:millisecondmagnetars}).

For many topics touched upon here, we can refer the interested reader to existing, excellent reviews; on GRB blastwaves \cite{Meszaros2012,2007PhR...442..166N,Waxman00,Meszaros2002,Piran2005}; on short and long GRB observations \cite{2007PhR...442..166N,2009ARA&A..47..567G}; on models of central engines involving core-collapse events \cite{Ott2008}, neutron star mergers \cite{2007NJPh....9...17L,Lattimer:2006}, and isolated neutron stars \cite{2009astro2010S.229O,corsi09}; on GW emission from core-collapse events \cite{Ott2008,2003LRR.....6....2N} and binary  mergers \cite{Cutler:2001,2002grg..conf...72C,2009CQGra..26k4004F,2010CQGra..27k4002D,lrr-2011-6};
on short GRB rates \cite{2007PhR...442..166N}, long GRB rates \cite{Kalogera:2004tn,Kalogera:2004nt}, and  compact object merger rates \cite{2010CQGra..27q3001A}.
Some investigations of isotropic prompt and delayed electromagnetic counterparts have recently been summarized in \cite{2012ApJ...746...48M} (for short GRBs).

This review is organized as follows. Section \ref{section:emissionprocesses} outlines GW emission processes from GRB progenitors, focusing on the emission scenarios that may produce signals which could be detectable with advanced GW detectors. Section \ref{section:astrophysics} presents some of the intriguing astrophysical questions that could be answered through the detection of GWs from GRB progenitors, in some cases in coincidence with other messengers. In Section \ref{section:observationalstrategies} we review GW observation strategies from single GW to multimessenger searches, also outlining prospects for the advanced detector era. Finally, Section \ref{section:conclusion} we briefly summarizes the presented results and the near future of GRB astrophysics with GWs.

\section{Gravitational-wave emission processes in GRB progenitors}
\label{section:emissionprocesses}

In this section, we outline the main GRB progenitor models with an emphasis on their expected GW signature. It should be clear that, depending on the progenitor model, the GW emission from GRB progenitors varies widely with respect to predicted signal strength and characteristic frequency.

\subsection{Compact binary coalescence}

The majority of short-hard GRBs are thought to be powered by the merger of NS-NS or NS-BH
binaries~\cite{1984SvAL...10..177B,1986ApJ...308L..43P,1989Natur.340..126E,2007NJPh....9...17L,2010CQGra..27k4002D,2012arXiv1204.4919Z,2008arXiv0809.1602K,2012arXiv1210.8152G}.
Compact binaries are also strong GW sources: the GWs produced in the minutes before the merger should be detectable, by second
generation instruments (i.e., Advanced LIGO and Virgo), reliably out to $\sim 450\,$Mpc for NS-NS
binaries, and further for NS-BH systems \cite{2010CQGra..27q3001A,2010ApJ...716..615O}.

\begin{figure}
\begin{center}
\resizebox{1\textwidth}{!}{\includegraphics{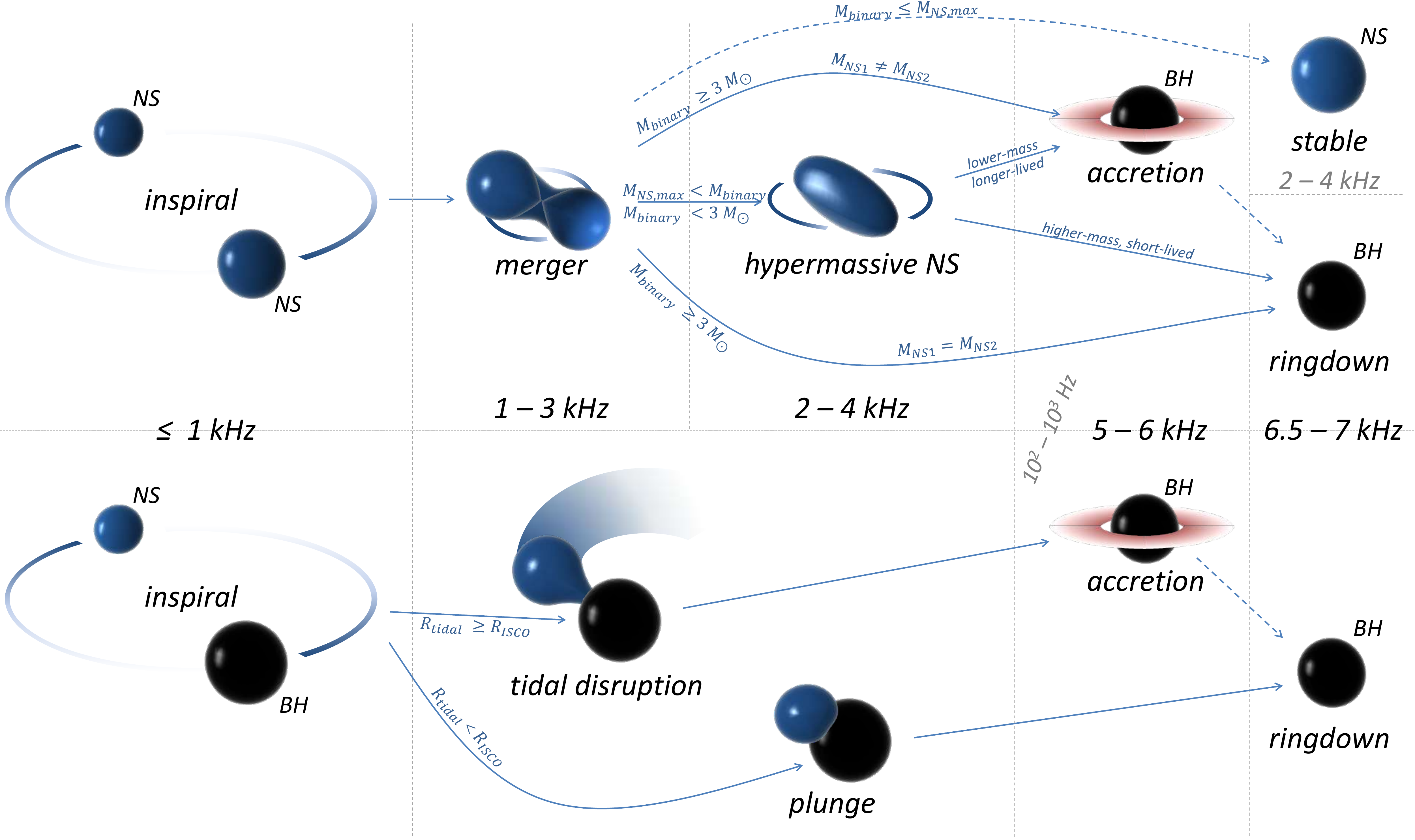}}
\end{center}
\caption{Schematic diagram of the evolution of compact binary coalescences. The frequency of the emitted GW is indicated for the different stages. \textbf{NS-NS} inspirals are observable for a few seconds to minutes. Upon the merger of the NSs, a binary with total mass $M_{binary}\gtrsim3\,$M$_\odot$ promptly collapses into a BH. For non-equal-mass binaries, the forming BH will be surrounded by an accretion disk. NS-NS binaries with total mass $M_{NS,max}<M_{binary}<3\,$M$_\odot$ (where $M_{NS,max}$ is the mass limit of non-rotating NSs) form a hypermassive NS with strong differential rotation, which assumes a non-axisymmetric ellipsoid shape. The hypermassive NS survives for milliseconds to a second, eventually collapsing into a BH, potentially with an accretion disk. Very low mass NS-NS binaries ($M_{binary}<M_{NS,max}$) can leave a stable NS behind. For \textbf{BH-NS} binaries, after an inspiral phase observable for seconds to minutes, the NS either gets tidally disrupted (if tidal disruption at radius $R_{tidal}$ occurs before the NS could reach the ISCO at $R_{ISCO}$), or it plunges into the BH (if $R_{tidal}<R_{ISCO}$). Tidal disruption results in a BH with an accretion disk, while no accretion disk forms upon plunge. This merger phase, along with the ringdown of the BH after plunge, lasts for milliseconds.} \label{fig:CBCflowchart}
\end{figure}

\begin{figure}
\begin{center}
\resizebox{1\textwidth}{!}{\includegraphics{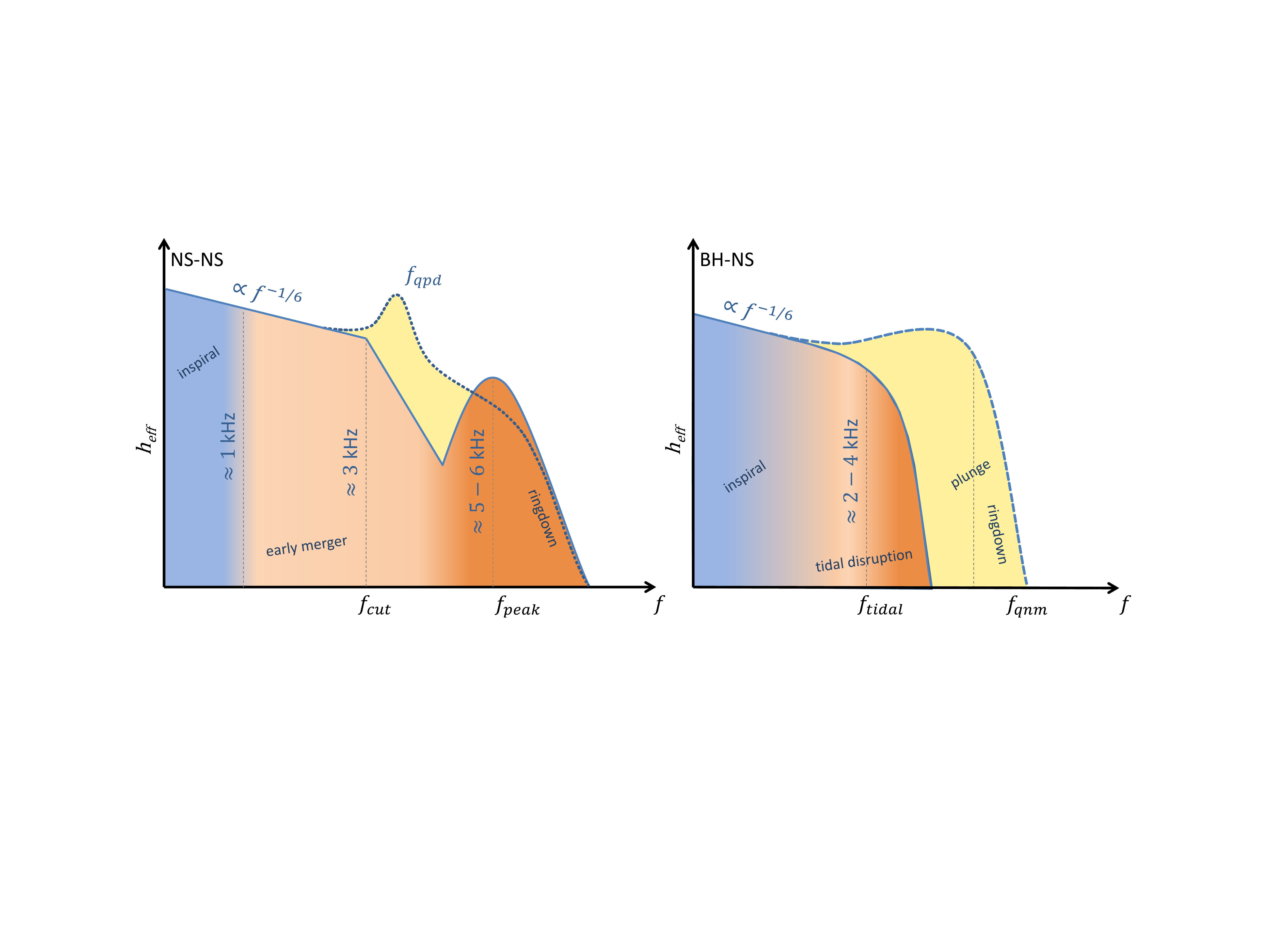}}
\end{center}
\caption{Schematic spectrum of effective amplitude $h_{\mathrm{eff}}$ for compact binary coalescences. \textbf{(left)} NS-NS binaries. During the inspiral phase up to $\approx1\,$kHz and the early-merger phase up to $f_{\mathrm{cut}}\sim3\,$kHz, the system retains its binary-like structure and $h_{\mathrm{eff}}$ scales as $f^{-1/6}$. If a BH is promptly formed, matter quickly falls in the BH, losing angular momentum through emitting GWs around a peak frequency $f_{\mathrm{peak}}\sim5-6\,$kHz. If a protoneutron star is formed from a NS-NS binary, it will radiate GWs through its quasiperiodic rotation at $f_{\mathrm{qpd}}\sim2-4\,$kHz. After matter falls into the BH, the BH rings down, emitting GWs at $\approx 6.5-7\,$kHz with exponentially decaying amplitude. \textbf{(right)} BH-NS binaries. During the inspiral phase, $h_{\mathrm{eff}}$ scales as $f^{-1/6}$. If the NS is tidally disrupted before reaching the ISCO, GW emission will cut off at $f_{\mathrm{tidal}}\sim2-4\,$kHz, i.e. the GW frequency at tidal disruption \cite{2011PhRvD..84f4018K,2012PhRvD..85d4061L}. If the NS plunges into the BH without being tidally disrupted, the plunge cuts off GW emission from the binary and excites the quasinormal mode of the remaining BH, which rings down emitting GWs at frequency $f_{\mathrm{qnm}}$ (NS-NS representation was partially inspired by Kiuchi et al. \cite{2010PhRvL.104n1101K}; BH-NS representation is based on Kyutoku et al. \cite{2011PhRvD..84f4018K}).}
\label{fig:CBCspectrum}
\end{figure}

Theoretical studies provide detailed information about the binary evolution and the gravitational waveform expected from these systems. Moreover, numerical simulations, which include the physics of the NS matter, continue to improve (see Refs. \cite{Lattimer:2006,2009CQGra..26k4004F,2010CQGra..27k4002D,lrr-2011-6} for reviews).
Here we present an overview of compact binary evolution, which is summarised in Fig.~\ref{fig:CBCflowchart}. At the highest level, there can be four phases: the inspiral phase during which the orbit shrinks due to energy and angular momentum loss via GWs; the merger phase during which dynamical effects and microphysics of the NS become important; the accretion phase when material stripped from the NS(s) accretes onto the merged objects; and the ringdown phase when the merged objects settle down to an unperturbed (Kerr) BH or a NS. The details vary by NS-NS and NS-BH merger and by mass (and other parameters) of the merging objects. The accretion stage is likely essential to the formation of short GRBs from binary coalescences, although the formation and evolution of protomagnetars following NS-NS merger has also been proposed as a plausible central engine \cite{2008MNRAS.385.1455M}.
Figure~\ref{fig:CBCspectrum} presents the main features of GW emission in a schematic spectrum of the GW effective amplitude, separately for NS-NS and BH-NS mergers. Below we discuss the different stages of the evolution in more detail.

\subsubsection{Inspiral phase}

--- Early in the evolution of a compact binary system, the two objects are separated by a relatively large distance compared to their radii. The binary elements spiral towards each other by losing angular momentum via the emission of GWs. During the early stages of the inspiral, the two compact objects can be approximated as point masses for the purposes of GW emission~\cite{lrr-2011-6}. At the late stages of the inspiral, the internal structure of the objects become increasingly important. For example, the tidal deformation of NSs in a binary system can (slightly) affect the orbital period (and therefore the gravitational waveform) in the late inspiral phase \cite{2008PhRvD..77b1502F,2009PhRvD..79l4033R,2009PhRvD..80f4037K,2010PhRvL.105z1101B,2011PhRvD..84b4017B}. Further, general relativistic spin-spin or spin-orbit coupling can cause the binary's orbital plane to precess, affecting the binary's evolution and GW emission \cite{1994PhRvD..49.6274A,2004PhRvD..70d2001V,2006PhRvD..74l2001L}.

Nevertheless, the dominant features of the GW signal from the inspiral phase are captured by neglecting the spins and internal structure of the binary elements. As the objects spiral together, their orbital frequency increases producing a GW signal that sweeps upward in frequency. About $\sim15$ minutes before merger, the GW from the inspiral of a NS-NS binary begins to sweep upward from $\sim10\,$Hz through the band of Earth-based GW interferometers. The effective amplitude $h_{\mathrm{eff}} \equiv f | \tilde{h}(f) |$ of the GW signal from a binary system decreases as $h_{\mathrm{eff}} \propto f^{-1/6}$ \cite{2010PhRvL.104n1101K}, up to a mass-dependent cutoff frequency $f_{\mathrm{cut}}\sim1-3\,$kHz \cite{1998PhRvD..57.4535F,2007PhRvL..99l1102O,2010PhRvL.104n1101K}. The frequency ranges $\lesssim1\,$kHz and $1-3\,$kHz are traditionally considered the inspiral and early merger phases, respectively. For $f\lesssim f_{\mathrm{cut}}$ the merger retains a binary-like structure and consequently emits relatively strong GWs \cite{2010PhRvL.104n1101K}.

Advanced detectors will be able to detect a NS-NS inspiral up to $D_h \sim 450\,$Mpc, while NS-BH inspirals will be detectable up to $D_h \sim950\,$Mpc~\cite{2010CQGra..27q3001A} (the distances are given for untriggered searches, with optimal source orientation and direction; for further details see Section \ref{section:allsky}). The effective survey volume determined by averaging over sky location and inclination of the sources is $\sim 4\pi (D_h / 2.26)^3/3$ \cite{2010CQGra..27q3001A}. Using the current best-guess rates of mergers, this gives tens of NS-NS and a few NS-BH binaries detected with advanced detectors each year~\cite{2010CQGra..27q3001A}. Additional advanced detectors, such as KAGRA \cite{2011IJMPD..20.1755K} or LIGO India \cite{2012CQGra..29l4012W}, can significantly increase this range \cite{2012arXiv1210.6362N}. Third generation detectors are expected to reach an order of magnitude farther than advanced detectors, i.e. to several Gpc, and hence will be able to observe tens of thousands of events a year (e.g., \cite{2011GReGr..43..409A}).

\subsubsection{Merger phase}
\label{subsubsection:mergerphase}

--- Depending on the binary system, the merger can progress in multiple distinct directions with qualitatively different GW and gamma-ray emission. The formation of a massive accretion disk is probably crucial to the generation of a GRB from binary mergers\footnote{Millisecond magnetars represent plausible alternative central engine candidates for some GRBs \cite{2007RMxAC..27...80T,2008MNRAS.385.1455M,2009MNRAS.396.2038B,2012MNRAS.419.1537B}. See also, e.g., \cite{ADM:ME2002} and references therein, for further alternative models.} \cite{2010CQGra..27k4002D}. Accretion disks can be created via the tidal disruption of a NS at some point during the merger \cite{2004MNRAS.351.1121R,2006PhRvD..73f4027S,2009PhRvD..79d4030S,2009PhRvD..79d4024E,2010PhRvD..82d4049K,2010PhRvD..81f4026F}. Alternatively, for NS-NS mergers, material with centrifugal support can be left outside the newly formed BH. Whether an accretion disk is formed or not depends on the binary properties (mass, spin, etc.), as well as the NS equation of state (EOS).
A disk mass of $\sim0.01\,$M$_\odot$, where M$_\odot$ is the mass of the sun, is probably sufficient to supply the energy for the creation of a short GRB \cite{2010PhRvL.104n1101K,2012arXiv1210.8152G}.
In a recent comparison between numerical simulations and observation, Giacomazzo et al. \cite{2012arXiv1210.8152G} found that the observed emission of short GRBs implies torus masses $\lesssim0.01\,$M$_\odot$, which favors ``high-mass" NS-NS mergers ($M_{total}\gtrsim3\,$M$_\odot$). Further, they find that BH-NS mergers, while cannot be excluded, would require a very rapidly spinning BH (with spin $\gtrsim0.9$).

Here we list the possible outcomes of the merger and outline the scenarios that can lead to them.

\begin{enumerate}[$\bullet$]
\item {\bf BH with no accretion disk} --- for BH-NS binaries, if the radius of the innermost stable circular orbit (ISCO) is greater than the tidal disruption radius, the NS plunges into the BH before it could be tidally disrupted, resulting in a BH with no accretion disk. This will be the case for binaries with relatively high BH:NS mass ratio ($\gtrsim5:1$) \cite{2009PhRvD..79d4030S,2009PhRvD..79d4024E,2010CQGra..27k4002D,2010PhRvL.105k1101C}. This ratio strongly depends on the BH spin (due to the change in the location of the ISCO) and the NS EOS \cite{2010PhRvD..82d4049K,2011PhRvD..84f4018K,2011ApJ...727...95P,2012arXiv1207.6304F}.

    For NS-NS binaries with equal masses, if the binary mass exceeds a threshold $M_{\mathrm{thr}}$, the NSs will promptly collapse to a BH upon merger \cite{2008PhRvD..78h4033B,2011PhRvD..83l4008H}, leaving essentially no accretion disk behind \cite{2006PhRvD..73f4027S,2010CQGra..27k4105R,2011PhRvD..83l4008H}. $M_{\mathrm{thr}}$ depends on the NS EOS. Kiuchi et al. \cite{2009PhRvD..80f4037K} found that a BH is promptly formed if total mass of the binary system is $\gtrsim 3\,$M$_\odot$.

    In this scenario no accretion-powered GRB will be produced, although there are possible channels through which even such a system can emit electromagnetic radiation \cite{2011ApJ...742...90M,2012ApJ...755...80P}.

\item {\bf BH with accretion disk} --- For BH-NS binaries with relatively low mass ratio ($\lesssim4:1$), the NS will be tidally disrupted before falling into the BH, which leads to the formation of a massive accretion disk \cite{2008PhRvD..77h4015S,2009PhRvD..79d4030S,2009PhRvD..79d4024E,2010CQGra..27k4002D}. For spinning BHs, the mass ratio below which disruption occurs is even higher \cite{2010PhRvD..82d4049K,2010PhRvL.105k1101C,2010CQGra..27k4106D,2011PhRvD..84f4018K}, and the mass of the formed disk can greatly depend on BH spin and spin-alignment \cite{2010PhRvD..82d4049K,2011PhRvD..83b4005F,2011PhRvD..84f4018K,2012arXiv1207.6304F}. Prior to tidal disruption, the emitted GW frequency reaches $f_{\mathrm{tidal}}\sim2-4\,$kHz\footnote{Mass shedding commences much earlier, at significantly GW frequencies than tidal disruption. Note that Ref. \cite{2000PhRvL..84.3519V} identified mass shedding as the cutoff frequency, which underestimates the tidal disruption frequency \cite{2012PhRvD..85d4061L}.} \cite{2011PhRvD..84f4018K,2012PhRvD..85d4061L}.

    For NS-NS binaries with unequal NS masses and sufficiently large total mass ($\gtrsim 3\,$M$_\odot$ \cite{2009PhRvD..80f4037K,2010CQGra..27k4105R,2011PhRvL.107e1102S}), the less massive NS will be tidally disrupted, followed by the more massive NS's prompt collapse into a BH, leaving a potentially massive accretion disk behind \cite{2008PhRvD..78h4033B}.

    In this scenario, after the merger of a NS-NS binary, if the forming disk features azimuthal variations, GWs may be emitted by the material orbiting the central object around a peak amplitude $f_{\mathrm{peak}}\sim5-6\,$kHz \cite{2010PhRvL.104n1101K}. For BH-NS binaries, $f_{\mathrm{peak}}$ can be significantly lower, $\sim$inversely proportional to the total mass of the system \cite{2009CQGra..26a5009H}. For instance a BH-NS binary with $5\,$M$_\odot$ BH mass has $f_{\mathrm{peak}}\approx 1-2\,$kHz, depending on, e.g., the BH spin \cite{2010PhRvD..82d4049K,2011PhRvD..84f4018K}.

    This scenario is a good candidate for the creation of accretion-powered GRBs.

\item {\bf Hypermassive NS formation} --- NS-NS binaries with total mass $\lesssim3$M$_\odot$ (this mass threshold depends on the NS EOS \cite{2009PhRvD..80f4037K,2011PhRvD..83l4008H,2011PhRvL.107e1102S}) will not promptly collapse into a BH, but will first form a so-called \emph{hypermassive} NS \footnote{Neutron stars are called hypermassive if they exceed the mass limit of rigidly rotating NSs \cite{2000ApJ...528L..29B}. NSs that are above the mass limit of non-rotating NSs, but whose mass could be supported by rigid rotation, are called \emph{supramassive}; e.g., \cite{2006PhRvL..96c1101D}. For a binary with low-mass ($\sim1\,$M$_\odot$) NSs, it is possible that the resulting post-merger NS mass can be supported even without differential rotation (i.e. it is not hypermassive). In this case the NS can be long lived (see, e.g., \cite{2010ApJ...724L.199O}).} \cite{2005PhRvD..71h4021S,2006PhRvD..73f4027S,2008PhRvD..78h4033B}, supported by differential rotation and thermal pressure \cite{2010CQGra..27k4002D,2010PhRvL.104n1101K,2011PhRvL.107e1102S}. The hypermassive NS eventually collapses into a BH with a delay of $1\,$ms-1\,s due to (i) losing angular momentum via GWs or magnetic processes \cite{2011PhRvD..83d4014G} (magnetorotational instability \cite{2006PhRvL..96c1101D} or magnetic winding \cite{2011ApJ...732L...6R}), (ii) its rotation becoming more uniform due to magnetic braking and viscosity \cite{2000ApJ...544..397S,2004PhRvD..69j4030D}, in which case differential rotational no longer supports the NS, and/or (iii) cooling due to, e.g., neutrino emission, so thermal pressure provides less support against the gravitational pull \cite{2011PhRvL.107e1102S,2011PhRvL.107u1101S}.

    A rapidly rotating hypermassive NS will assume a (non-axisymmetric) ellipsoidal shape \cite{2008PhRvD..78h4033B}, which is energetically favorable over a spheroid \cite{2010CQGra..27k4002D}. Such an ellipsoidal hypermassive NS will emit a strong GW signal at twice its (quasiperiodic) rotational frequency $f_{\mathrm{qpd}}\sim2-4\,$kHz \cite{2007PhRvL..99l1102O,2009PhRvD..80f4037K,2011PhRvD..83l4008H,2011PhRvL.107e1102S}.

    Hypermassive NSs may leave a massive accretion disk behind after collapsing into a BH \cite{2006PhRvD..73f4027S,2011PhRvD..83l4008H,2012arXiv1204.6240R}. It seems that the outcome depends on the binary mass. As we saw above, for high-mass NSs, the merger promptly forms a BH. At the highest masses for which a hypermassive NS is formed, a very short-lived NS is formed, with likely suppressed accretion disk formation \cite{2011PhRvD..83l4008H}. For lower-mass binaries, the formed hypermassive NS is longer-lived, resulting in stronger GW emission. These lower-mass binaries also result in the formation of a more massive accretion disk.

    In this scenario, in which a hypermassive NS is formed, a significant amount of GW energy can be emitted from the hypermassive NS at around its quasiperiodic rotation frequency. Quasiperiodic GW emission from a hypermassive NS would be detectable with advanced detectors from $\sim20\,$Mpc \cite{2005PhRvL..94t1101S,2011PhRvL.107e1102S}, especially because it would be accompanied by an inspiral phase with significantly higher signal-to-noise ratio (SNR). This scenario is a good candidate for the creation of GRBs, with either BH-torus or protomagnetar central engines.

\item {\bf Formation of stable NS} --- If the total mass $M_{binary}$ of a NS-NS binary is below the mass limit $M_{NS,max}$ of non-rotating NSs, or if the merged NS's mass is $<M_{NS,max}$ due to, e.g., tidal disruption, a stable, long-lived NS can form from the merger. While typical observed NS-NS binary masses \cite{1999ApJ...512..288T} are likely above $M_{NS,max}\gtrsim2\,$M$_\odot$ \cite{2010Natur.467.1081D}, it is plausible that some mergers end up forming a stable NS.
\end{enumerate}

Accretion disks themselves may contribute to the GW emission of binaries \cite{2003MNRAS.341..832Z,2005MNRAS.356.1371Z,2007PhRvD..75d4016N,2011PhRvL.106y1102K}. BH-torus systems can be unstable to non-axisymmetric perturbations that may give rise to non-axisymmetric torus structure, resulting in GW emission at a few-hundred Hz. See also Section \ref{section:accretiondiskinstabilities} for GW emission from accretion disks.

\subsubsection{Ringdown phase}

--- When a BH is formed, or when substantial matter with non-axisymmetric structure (i.e., NS or fragmented accretion disk) falls into the BH, the event horizon of the BH is initially perturbed. Subsequently, it quickly approaches the non-perturbed (Kerr) solution via the emission of GWs. This process is called ringdown, with characteristic frequencies of $f_{\mathrm{peak}}\approx 6.5-7\,$kHz  (e.g., \cite{2009PhRvD..80f4037K,2011PhRvD..83l4008H}) \footnote{Compare to the orbiting frequency $f_{\mathrm{ISCO}}=4.4(M/M_\odot)^{-1}\,$kHz of a particle around a Schwarzschild BH of mass $M$ at the innermost stable circular orbit (ISCO). The orbital frequency greatly depends on the BH spin (e.g., \cite{2012PhRvD..85f2002R} for the specific dependence).}, given the typical mass of the binary ($\sim2-3\,$M$_\odot$).

At and above $f_{\mathrm{peak}}$, the GW spectrum is qualitatively independent of the properties of the binary (although quantitatively the characteristic frequency decreases with increasing binary mass). It decays exponentially due to the $m=2$ quasinormal mode oscillation of the final black hole \cite{1998PhRvD..57.4535F,2010CQGra..27k4002D}. For solar-mass binaries, GW radiation from the BH ringdown is undetectable with advanced GW interferometers due to its high frequency.

We note that BH ringdown can be suppressed or even choked in the presence of intense accretion onto the BH, as is likely the case for unequal-mass NS-NS binaries \cite{2010CQGra..27k4105R}.

\subsection{Core Collapse}
\label{section:collapsars}

Massive stars develop an inert iron core supported by non-thermal degeneracy pressure in their center as the final stage of nuclear fusion. The growing iron core becomes gravitationally unstable upon reaching a mass around the Chandrasekhar mass ($\sim1.44\,$M$_\odot$) due to a softening of the EOS as the degenerate fermions become relativistic \cite{2007PhR...442...38J}, collapsing into a NS or a BH. Such a collapsing stellar system may result in a supernova explosion (a core-collapse supernova), and/or a GRB. The central engine driving GRB emission may be an accreting BH (e.g., \cite{Wo:93}) or a rapidly spinning, strongly magnetized protoneutron star (a \emph{millisecond proto-magnetar}; e.g., \cite{2011MNRAS.413.2031M} and references therein).

\begin{figure}
\begin{center}
\resizebox{1\textwidth}{!}{\includegraphics{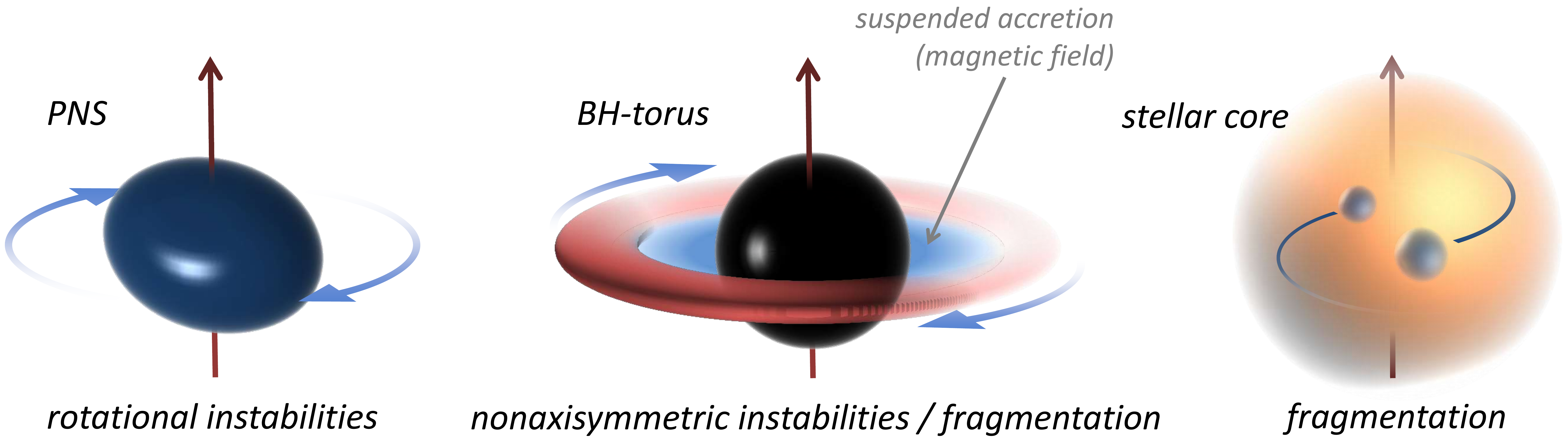}}
\end{center}
\caption{Schematic diagram of GW emission scenarios of massive stellar core collapses.} \label{fig:collapsarGW}
\end{figure}

The core collapse of a massive star may emit GWs through various mechanisms \cite{2003LRR.....6....2N,Ott2008}.
Below we focus on the mechanisms producing GWs that may be sufficiently strong to be detected at distances relevant to GRBs, i.e. that may be detectable from $\gg10\,$Mpc with 2$^{nd}$ or $3^{rd}$ generation GW interferometers. A schematic diagram of these emission processes are shown in Fig. \ref{fig:collapsarGW}.

\subsubsection{Rotational instabilities in protoneutron stars}
\label{section:protoneutronstar}

--- For massive stars with initial stellar masses $10\,$M$_\odot\lesssim M \lesssim25\,$M$_\odot$, the collapsing core is expected to form a so-called protoneutron star \cite{2003ApJ...591..288H}. The resulting rotating protoneutron star can be unstable to non-axisymmetric deformations, potentially giving rise to copious GW emission \cite{1998LRR.....1....8S,2000ApJ...542..453S,2002ApJ...565..430F,2003CQGra..20R.105A,2007PhRvD..75d4023B,2007CQGra..24S.171M,2010CQGra..27k4104C}. The onset of rotational instabilities depends on the rotational rate of the star, which can be conveniently parameterized by $\beta \equiv T_{\mathrm{rot}}/|W|$, i.e. the ratio of the star's rotational kinetic ($T_{\mathrm{rot}}$) and gravitational potential ($W$) energy \cite{1994PhRvL..72.1314H}. The resulting non-axisymmetric structure may be a bar-like $m=2$ mode, giving rise to a characteristic GW emission. Higher $m$ modes may also arise, albeit they have longer growth time \cite{1983PhRvL..51...11F}.
While there are still various uncertainties in the evolution and role of rotational instabilities (e.g., non-linear mode-coupling effects can severely limit the deformation \cite{2007PhRvD..75d4023B}, or the role of viscosity and neutrino cooling), the emerging picture is that rotational instabilities are likely viable emitters of GWs and could play an important role in the future detection and understanding of GRB progenitors through their GW signature.

The energy potentially available for GW emission is abundant. The rotational energy of a typical NS with 1$\,$kHz rotational frequency is $\sim10^{-2}$M$_\odot\,$c$^2$ (e.g., \cite{2001ApJ...550..426L}). Even a fraction of this energy, if radiated away in GWs, could be detectable at large distances ($\gg10\,$Mpc) with advanced detectors. The protoneutron star may also accrete supernova fallback material. Such an accretion further increases the angular momentum and energy available that may be radiated away via GWs \cite{2002MNRAS.333..943W,2008arXiv0809.1602K,2012arXiv1207.3805P}.

The amplitude of a GW signal emitted by a rotating bar scales as $h\sim M R^2 f^2 /d$, where $M$, $R$, $f$ and $d$ are the mass, radius, GW frequency (i.e. twice the rotational frequency), and distance of the NS, respectively \cite{2008arXiv0809.1602K}. The energy radiated away in GWs, in the Newtonian quadrupole approximation, can be estimated as (see, e.g., \cite{2008arXiv0809.1602K})

\begin{equation}
E_{\mathrm{GW}} \approx 10^{-2} M_\odot c^2 \left(\frac{\epsilon}{0.2}\right)^2 \left(\frac{f}{2\,\mbox{kHz}}\right)^6\left(\frac{M}{1.4\,M_{\odot}}\right)\left(\frac{R}{12\,\mbox{km}}\right)^2\left(\frac{\tau}{0.1\,\mbox{s}}\right),
\label{equation:barmodeEGW}
\end{equation}
where $\epsilon$ is the ellipticity of the bar and $\tau$ is the duration of the presence of the instability.
\newline
Protoneutron stars are subject to different rotational instabilities, the two main categories being dynamical and secular instabilities. A star is dynamically unstable if it is unstable to non-axisymmetric perturbations even in the absence of dissipation (i.e. if a slightly non-axisymmetric shape that conserves angular momentum is energetically favorable). A star is secularly unstable if it is unstable to non-axisymmetric perturbations only if dissipative effects are relevant, i.e. if the change towards a non-axisymmetric shape requires the radiation or redistribution of angular momentum (see \cite{CPA:CPA3160200203,Lebovitz19981407} for interesting historical overviews). Below we discuss these instabilities further in detail.
\begin{enumerate}[$\bullet$]
\item {\bf Dynamical instabilities} -- Rapidly rotating stars will be subject to dynamical instabilities driven by hydrodynamical and gravitational effects \cite{1998LRR.....1....8S,2005ApJ...618L..37W,2007PhRvD..75d4023B}. Dynamical instability is the ``simplest" form of NS instabilities, since its development is quick (on the time scale of the rotational period of the NS), and does not require dissipation. A uniformly rotating, classical fluid body becomes dynamically unstable at a rotation rate $\beta \gtrsim0.27 \equiv \beta_{\mathrm{dyn}}$ \footnote{Note that a uniformly rotating, constant density star cannot reach such high $\beta$. The maximum rotation of a compact star is limited by mass-shedding to $\beta\sim0.1$ \cite{2003CQGra..20R.105A}.} \cite{Chandrasekhar:207046}. Stability conditions are essentially the same for relativistic stars, for which $\beta_{\mathrm{dyn}}\sim0.24$ \cite{2007PhRvD..75d4023B,2008arXiv0809.1602K}. Furthermore, differentially rotating stars are subject to non-axisymmetric instabilities even at much slower rotation with $\beta\lesssim0.09$ \cite{2005PhRvD..71b4014S,2007PhRvL..98z1101O,2007CQGra..24..139O,2008A&A...490..231S}. Such low-$\beta$ instabilities in differentially rotating protoneutron stars are probably analogous to the Papaloizou-Pringle instability (\cite{1984MNRAS.208..721P}; also see Section \ref{section:accretiondiskinstabilities}):  the protoneutron star core is surrounded by a fluid rotating at the frequency of a non-axisymmetric mode of the core, hence exciting this mode \cite{2005ApJ...618L..37W}. As numerical simulations so far have been too short to capture the long-term behavior of some dynamical instabilities, they are not conclusive in terms of the total energy emitted via GWs from dynamical instabilities. Nevertheless, GW emission seems to be fast relative to the cooling time of the protoneutron star, or compared to energy loss due to viscosity. Consequently, if competing mechanisms that radiate away angular momentum (e.g., magnetic fields) are weak, GWs can carry away a significant fraction of the protoneutron star's rotational energy, producing a signal that may be detectable from
    \begin{equation}
    D\gtrsim 60\,\mbox{Mpc}\left(\frac{E_{\mathrm{GW}}}{10^{-2}\,\mbox{M}_\odot\mbox{c}^2}\right)^{1/2}\left(\frac{f}{1\,\mbox{kHz}}\right)^{-1}
    \end{equation}
    for narrow-band, circularly polarized GW signals from optimal source direction\footnote{Given $10\times$ sensitivity improvement for advanced detectors; based on the results of \cite{2008PhRvL.101u1102A} that used only one GW detector -- multiple detectors could see even farther.}. The frequency scaling of the distance only applies to $f\gtrsim300\,$Hz.

\item{\bf Secular instabilities} -- Stars with lower rotation rates (i.e. for which the faster-developing dynamical instabilities are not prevalent) can be subject to secular (i.e. dissipation-driven) non-axisymmetric instabilities \cite{1995ApJ...442..259L,1998PhRvL..80.4843L,2009ApJ...702.1171C}. Dissipation can occur via gravitational radiation \cite{1998PhRvL..80.4843L} or fluid viscosity \cite{1998LRR.....1....8S,2008arXiv0809.1602K}. GW emission drives frame-dragging (so-called Chandrasekhar-Friedman-Schutz) instabilities of modes that are retrograde with respect to the star but prograde with respect to the observer \cite{1970PhRvL..24..611C,1978ApJ...222..281F}. Among GW-driven instabilities, fundamental $f$-mode bar instabilities have the shortest growth time: $0.1\,\mbox{s} \lesssim \tau_{\mathrm{GW}}\lesssim7\times10^4\,$s for $0.27\gtrsim\beta\gtrsim0.15$ \cite{1995ApJ...442..259L}. In the uniformly rotating approximation of a relativistic star, the protoneutron star becomes unstable to GW-driven $m=2$ $f$-mode instabilities for $\beta\gtrsim0.06-0.09\equiv \beta_{\mathrm{sec}}$ depending on the EOS and stellar mass \cite{1999ApJ...510..854M} (compare with the Newtonian limit of $\beta\gtrsim0.14$ \cite{1970PhRvL..24..611C,1995ApJ...442..259L}). At lower rotation rates, stars become unstable to higher-multipole $f$ modes, albeit higher modes have longer growth times \cite{1983PhRvL..51...11F,2012arXiv1209.5308P}. A recent work of Passamonti et al. \cite{2012arXiv1209.5308P} indicates that the most unstable $f$ modes developing in the aftermath of a supernova explosion may in fact be $l=m=3$ and 4 modes.

    As angular momentum is radiated away through GWs (and potentially other channels), the protoneutron star's rotation frequency and therefore the emitted GW frequency decreases, sweeping towards the most sensitive band of LIGO-like detectors. Some analytical and numerical results indicate that such a GW signal may be detectable at large distances, up to $\sim100\,$Mpc with advanced interferometers \cite{1995ApJ...442..259L,2004PhRvD..70h4022S,2004ApJ...617..490O,2010PhRvD..81h4055Z}. Some other recent simulations are also promising, even though they only cover the first few milliseconds after core collapse and therefore cannot capture the long-term evolution of the bar mode \cite{2007PhRvL..98z1101O,2007CQGra..24..139O}. Nevertheless, some recent, realistic simulations (e.g., \cite{2012arXiv1209.5308P,2012arXiv1206.6604C,2012PhRvD..85b4030Z,2012arXiv1203.3590L}) predict much smaller detectable range.

    Rotating protoneutron stars are also unstable to GW-driven \emph{r-mode} oscillations (the restoring force being the Coriolis force) at any (i.e. arbitrarily low) rotation rate \cite{1998ApJ...502..708A,2002CQGra..19.1247O,2008arXiv0809.1602K,2011PhRvL.107j1101H}. $R$ modes are important only if their growth time is less than the damping time of viscous forces. Further, $r$-mode instability is expected to be saturated at low amplitude due to dissipative effects \cite{2011GReGr..43..409A}, and is suppressed in the presence of magnetic fields \cite{2000ApJ...531L.139R}. Nevertheless, under favorable conditions, GW signals from protoneutron star $r$ modes may be detectable for several years after core collapse \cite{2009PhRvD..79j4003B}. Given such a long duration, the emitted GW signal may be integrated for a measurement of $\sim1\,$yr that would give a detectable signal to distances of $\sim30\,$Mpc \cite{2009PhRvD..79j4003B} (or to even farther if the protoneutron star is a \emph{strange quark star} \cite{2011GReGr..43..409A,2005PhRvL..95u1101O}). Further calculations (see \cite{2008arXiv0809.1602K} and references therein) suggest that the saturation (i.e. maximum) amplitude of $r$-mode instabilities is limited by its non-linear coupling to other inertial modes.

\item{\bf Magnetic distortion} -- Toroidal magnetic fields ($B\gtrsim10^{12}\,$G) can distort a rotating NS into a prolate shape \cite{2001MNRAS.327..639I,2002PhRvD..66h4025C,2003PhRvD..67l4026I,2004ApJ...600..296I,2012arXiv1207.4035F}. Such configuration is unstable to the growth of the angle between the NS's angular momentum and the magnetic axis. This angle increases until the angular momentum and magnetic axis are orthogonal. The resulting configuration is a rotating, non-axisymmetric body that is an efficient emitter of GWs \cite{2002PhRvD..66h4025C,2009MNRAS.398.1869D}. Fast rotating magnetars can also lose spin energy via magnetic dipole radiation and/or magnetized, relativistic winds. Nevertheless, observations indicate that such energy losses are not typical (\cite{2009MNRAS.398.1869D} and references therein), suggesting that GW emission may be relevant in the early phase of newly born magnetars. Given that magnetic distortions indeed result in efficient GW emission, Dall'Osso et al. \cite{2009MNRAS.398.1869D} find that such a GW signal can be detected out to the Virgo cluster with advanced detectors.
\end{enumerate}

Besides their initial rotation, protoneutron stars can accrete material from the infalling matter after core collapse \cite{1989ApJ...346..847C}. It is possible that NSs radiate away this angular momentum that they gain from accretion via GWs, e.g., via dynamical and/or secular instabilities discussed above. Such GW signal may be detectable to $\gtrsim10\,$Mpc with advanced detectors \cite{2012arXiv1207.3805P}.

\subsubsection{Accretion disk non-axisymmetric instabilities}
\label{section:accretiondiskinstabilities}

--- Upon the core collapse of a massive star ($\gtrsim 30\,$M$_{\odot}$ \cite{2006RPPh...69.2259M}), a plausible scenario is the formation of a central BH surrounded by an accretion disk \cite{1993ApJ...405..273W,2011ApJ...737....6S,2012arXiv1206.5927S}. Such a BH-torus system can be the source of copious GW emission if the disk assumes a finite quadrupole structure due to non-axisymmetric instabilities \cite{2003LRR.....6....2N}. The emergence of such a non-axisymmetric structure on the time scales comparable to the lifetime of the accretion disk requires a stellar progenitor with sufficiently high angular momentum. A high rotation rate is also required for the creation of a GRB (e.g., \cite{2005NatPh...1..147W}). Below we outline some of the possible scenarios through which non-axisymmetric instabilities in accretion disks may result in strong GW emission, and address whether existing or future GW detectors can probe that signal.

As an example to the sensitivity of a GW search to accretion-disk non-axisymmetric instabilities, we consider a 3\,M$_\odot$ BH with a $0.01\,$M$_\odot$ clump in its accretion disk. Such a system emits GWs similar to an isolated low-mass binary coalescence, and may be seen to a horizon distance (for definition see Section \ref{section:allsky}) \cite{2010ApJ...716..615O}
\begin{equation}
D_h\simeq 55\,\mbox{Mpc}\,(\mathcal{M}_c/0.1\,\mbox{M}_\odot)^{5/6}
\label{equation:horizonscaling}
\end{equation}
for an advanced detector with optimal source position and orientation (this distance is reduced by a factor 2.26 for average sky location and orientation), where $\mathcal{M}_c=(m_1m_2)^{3/5}/(m_1+m_2)^{1/5}$ is the so-called \emph{chirp mass} ($\mathcal{M}_c=0.1\,$M$_\odot$ for the example above).

\begin{enumerate}[$\bullet$]
\item {\bf Disk fragmentation via gravitational instability} -- Accretion disks with sufficiently large angular momenta are gravitationally unstable  \cite{1994ApJ...420..247W,2007ApJ...658.1173P,2004PhRvD..69j4016D}. A gravitationally unstable disk will fragment if the disk cooling time is sufficiently short ($\lesssim$ orbital period) \cite{2001ApJ...553..174G}. Relevant accretion disks cool rapidly, e.g., via neutrinos or strong winds \cite{2007NJPh....9...17L}, favoring fragmentation, although the emergence of magnetorotational instability (MRI) in the disk can heat the disk through ohmic dissipation \cite{2000ApJ...530..464F}. The resulting fragmented disk will emit a strong, chirp-like GW signal \cite{2003ApJ...589..861K}. Viscosity and GW emission drive the angular momentum loss of the disk. Consequently, low disk viscosity favors stronger GW emission, which may be detectable from $\sim100\,$Mpc with Advanced LIGO/Virgo \cite{2007ApJ...658.1173P}. The GW frequency at its highest SNR is probably in the $10^2-10^3\,$Hz range, depending on the strength of viscous forces. The duration of GW emission may be similar to that of gamma-ray emission. Furthermore, the tidal disruption of fragments in the accretion disk may be behind some of the X-ray flares observed in GRB afterglows \cite{2007ApJ...658.1173P}. Accretion disk fragmentation has been observed in approximate numerical simulations \cite{2004PhRvD..69j4016D,2011PhRvD..84b4022G}. It is not clear, however, whether such a fragmentation would occur in more realistic cases as well, unless the pressure support of the star is drastically reduced, e.g., via neutrino losses \cite{2011PhRvD..84b4022G}.

\item {\bf Non-axisymmetric structure via Papaloizou-Pringle instability} -- Differentially rotating accretion disks can be subject to global non-axisymmetric instabilities. One such instability, discovered by Papaloizou \& Pringle \cite{1984MNRAS.208..721P}, develops in accretion disks in which azimuthal pressure gradients (due to high internal temperature) give rise to differential rotation \cite{1984MNRAS.208..721P,1986MNRAS.221..339G,1996MNRAS.281..119C,1997ASPC..121...90B}, i.e. the angular velocity $\Omega(R)$ of the disk depends on the radius $R$. In such disks, large-scale spiral pressure waves can emerge with fixed pattern speed $\Omega_p$. If there is a radius $R_c$, the so-called corotation radius, at which the pattern speed is equal to the disk rotational rate (i.e. $\Omega(R_c)=\Omega_p$), the wave within $R_c$ will propagate slower than the rotation of the disk, in fact  decreasing the mechanical energy of the disk, resulting in a so-called \emph{negative energy wave} (e.g., \cite{1997ASPC..121...90B}). Such negative energy wave can interact, at the corotation radius, with the \emph{positive energy wave} that develops at $R>R_c$. The negative energy wave can increase its amplitude by \emph{losing} energy, which in turn can increase the amplitude of the positive energy wave, thus feeding the instability. See, e.g., \cite{1997ASPC..121...90B} for an expressive description of the phenomenon.

    The Papaloizou-Pringle instability gives rise to an ($m=1$) non-axisymmetric structure on a dynamical time scale (i.e. over a time period comparable to the rotation period). Such non-axisymmetric structure can persist for much longer than the dynamical time scale, resulting in strong GW emission. The Papaloizou-Pringle instability and the resulting non-axisymmetric structure have been observed in 3D relativistic simulations of BH-torus systems \cite{2011PhRvD..83d3007K,2011PhRvL.106y1102K}. These simulations indicate that BH-accretion disk systems subject to the Papaloizou-Pringle instability emit GWs in the $10^2-10^3\,$Hz frequency range that may be detectable up to $\sim100\,$Mpc with Advanced LIGO/Virgo \cite{2011PhRvL.106y1102K}.
    Strong magnetic fields present in the accretion disk can enhance the instability for thick disks (and may suppress it for thin disks) \cite{2011MNRAS.410.1617F}.
    Nevertheless, we note that the Papaloizou-Pringle instability was found in simulations for initially axisymmetric tori \cite{2011PhRvL.106y1102K}. Such complete axial symmetry may not develop in compact binary mergers. Numerical simulations of compact binary mergers have not yet shown signs of the development of the Papaloizou-Pringle instability (e.g., \cite{2011ApJ...732L...6R}).

\item {\bf Suspended accretion} -- In order for accretion around a central, rotating BH to continue for the time scales of long GRBs, it has been suggested that accretion may be ``suspended" (i.e. slowed down or temporarily halted; e.g., \cite{2001ApJ...552L..31V,2001ApJ...552L..31V,2003ApJ...584..937V}). Such suspended accretion would be achieved through magnetic fields, which transfer some of the rotational energy of the BH to the disk. A fraction of this rotational energy could then be radiated away through GWs \cite{2001PhRvL..87i1101V,2004PhRvD..69d4007V}. In order to emit GWs, the accretion disk needs to lose its axial symmetry, e.g., through fragmentation. GW radiation from suspended accretion has been suggested to carry away as much as $E_{\mathrm{GW}}\sim10^{-2}\,$M$_\odot$c$^2$ in the sensitive frequency band of LIGO/Virgo-like interferometers over a duration comparable to the duration of long GRBs ($\sim30\,$s) \cite{2001ApJ...552L..31V,2003ApJ...584..937V}. Suspended accretion, nevertheless, requires highly ordered accretion, which may be difficult to achieve due to the development of MRI heating or disk turbulence (e.g., \cite{2005ApJ...622.1008K}). Simulations to date (e.g., \cite{2011MNRAS.413.2031M}) have not provided support for suspended accretion, even in the presence of $\sim10^{15}$\,G magnetic fields.
\end{enumerate}

\subsubsection{Fragmentation of collapsing core}
\label{section:fragmentationofcollapsingcore}

--- In very rapidly rotating stars, infalling matter may fragment even before the formation of a BH-torus system. It has been suggested that, in analogy with observed phenomena in star formation, the collapsing core may fragment and form two or more compact objects \cite{1995MNRAS.273L..12B,2002ApJ...579L..63D,2003ApJ...589..861K}. Such fragmentation was observed in relativistic numerical simulations of approximate pre-supernova cores \cite{2006PhRvL..96p1101Z,2007PhRvD..76b4019Z,2006astro.ph..8028R}. Such core fragmentation would give rise to strong, characteristic GW emission, similarly to the case of binary mergers. Nevertheless, the rotation rate necessary for such a fragmentation seems to be difficult to achieve with current stellar models \cite{2003LRR.....6....2N}.

\subsection{Magnetars}
\label{subsection:magnetars}

Highly magnetized neutron stars (magnetars) are thought to be the engine behind soft gamma repeaters (SGRs) and anomalous X-ray pulsars (AXPs) \cite{1992ApJ...392L...9D,1995MNRAS.275..255T,1996ApJ...473..322T,1999ApJ...510L.115K}. Some magnetars occasionally produce so-called \emph{giant flares} that resemble short GRBs (\cite{2009ARA&A..47..567G} and references therein). Giant flares are much less energetic than typical short GRBs that are at cosmological distances, hence they are only detectable from the Milky Way or nearby galaxies, up to $\lesssim40\,$Mpc with current instruments \cite{2005Natur.434.1107P,2009ARA&A..47..567G}. The lack of excess short-GRB population from the direction of these galaxies indicates that magnetars may only be responsible for a small sub-population of short GRBs \cite{2005Natur.434.1107P}.

Giant flares are much more common (with much weaker gamma emission) than other types of GRBs, as indicated by the rate of observed galactic and nearby giant flares. Due to their high rate, they may be detectable with GW observatories even though they are much weaker sources of GWs than other, extragalactic GRBs.

Giant flares (as well as SGR activity or abrupt changes in the NS spin period; \cite{1996Natur.382..518C}) are thought to be the result of so-called \emph{starquakes}: the tectonic activity (cracking) of the NS crust \cite{1996Natur.382..518C}, which is followed by the reconfiguration of the NS's magnetic fields \cite{2005Natur.434.1107P}. Starquakes induce seismic vibrations in the NS, causing the observed quasi-periodic oscillations (QPOs) in the X-ray tails of giant flares \cite{1983A&A...126..400B,2005ApJ...632L.111S,2006ApJ...653..593S,2006ApJ...637L.117W,2005ApJ...628L..53I,2005ApJ...632L.111S}.

Alternatively, it has been suggested \cite{2006MNRAS.368L..35L,2011MNRAS.410.1036V} that QPOs cannot be driven by oscillations in the NS crust due to the quick ($\sim1\,$s) dissipation of the vibration via Alfv\'{e}n waves into the neutron star interior. Nevertheless, the similarities between the statistical properties of SGR events and earthquakes may provide further evidence for the crustal origin of SGR events \cite{2005Natur.434.1107P}.

NS seismic vibrations result in the emission of GWs \cite{1998MNRAS.299.1059A,2001MNRAS.320..307K,2001MNRAS.327..639I,2007PhRvD..76f2003A,2011PhRvD..83j4014C}, although this emission may be weak \cite{2005MPLA...20.2799H}. In the optimal scenario for GW emission, magnetic reconfiguration can liberate $10^{48}-10^{49}\,$erg of crustal elastic energy, which is the upper limit for the energy radiated away via GWs \cite{2011PhRvD..83j4014C,2001MNRAS.327..639I}. This energy can be even higher if the NS is of strange quark matter \cite{2011PhRvD..83j4014C}.

Quasi-periodic oscillations (QPOs) with various frequencies from $\sim10-10^3\,$Hz and durations up to $\sim100\,$s have been observed in every giant flare X-ray afterglow so far \cite{1983A&A...126..400B,2005ApJ...632L.111S,2006ApJ...653..593S,2006ApJ...637L.117W,2005ApJ...628L..53I,2005ApJ...632L.111S}. Some of these QPOs may be connected to the seismic modes of the NSs \cite{1983A&A...126..400B,2005ApJ...628L..53I}. Of special interest are QPOs around $\sim100\,$Hz, which fall into the most sensitive frequency band of LIGO/Virgo. Possible GW emission of $\sim10^{44}\,$erg at this frequency would be detectable out to $\sim10\,$kpc with Advanced LIGO/Virgo (given 10$\times$ sensitivity improvement compared to initial detectors) \cite{2008PhRvL.101u1102A}. It is possible that GWs are emitted for a significantly longer time (days to months) than the observed electromagnetic QPO \cite{2011PhRvD..83h1302K}, making longer-term GW searches necessary.

A fraction of the energy from magnetic reconfiguration may excite NS fundamental quadrupolar fluid modes (\emph{f modes}; e.g., \cite{2011ApJ...735L..20L,2011ApJ...736L...6C}, although see \cite{2012arXiv1206.6604C}). If NS $f$ modes are excited, they are damped by GWs on a very short time scale ($\sim200\,$ms; \cite{2011PhRvD..83j4014C} and references therein), which is shorter than most other potential damping mechanisms. Therefore, most of the energy in the $f$ modes may be radiated away via GWs \cite{2011PhRvD..83j4014C}. NS $f$ modes oscillate at $1-2\,$kHz (for stiff EOS \footnote{The density of NSs with stiff EOS changes relatively slowly with pressure (as opposed to soft EOS).}), not too far from the most sensitive band of Advanced LIGO/Virgo \cite{2001MNRAS.320..307K}. Given the most extreme case in which $10^{49}\,$erg being transferred into $f$ modes that oscillate at $\sim1\,$kHz, the resulting GW could be detected up to $\lesssim2\,$Mpc with Advanced LIGO/Virgo (given 10$\times$ sensitivity improvement compared to initial detectors) \cite{2008PhRvL.101u1102A}. ``Stacking" the GW signals from multiple events can extend this distance even farther \cite{2009ApJ...701L..68A,2011ApJ...734L..35A}. If the excitation energy is comparable to the observed giant flare energy ($\sim10^{46}\,$erg) and if not all energy is radiated away through GWs, only galactic sources would be detectable with advanced GW detectors. Furthermore, recent magnetohydrodynamic simulations of magnetized NSs indicate that only a small fraction of the released energy is converted to $f$ modes, which would make the detection of $f$-mode GWs less likely \cite{2012PhRvD..85b4030Z,2012arXiv1203.3590L}.

\subsection{Millisecond protomagnetars}
\label{section:millisecondmagnetars}

While the commonly favored model for the central engine behind GRBs is a rapidly accreting BH \cite{Wo:93,1999ApJ...526..152F,2001ApJ...557..949N,2010ApJ...713..800L}, there are observational indications \cite{2006Natur.444.1053G,2006Natur.444.1044G} that a fraction of short GRBs may originate from highly magnetized, rapidly spinning protoneutron stars, so called \emph{millisecond protomagnetars} \cite{2007RMxAC..27...80T,2008MNRAS.385.1455M,2009arXiv0908.1127M,2009MNRAS.396.2038B,2012MNRAS.419.1537B}. Further, outflows driven by such protomagnetars could also lead to the creation of long GRBs \cite{2004ApJ...611..380T,2007ApJ...659..561M,2011MNRAS.413.2031M}. The emergence of a millisecond protomagnetar from various cosmic events is therefore a plausible step in the creation of some GRBs.

The possibility that some GRBs are driven by rapidly spinning protomagnetars opens the door to alternative GRB progenitors. While traditionally considered progenitors, i.e. (i) the core collapse of massive stars \cite{2004ApJ...611..380T,2007ApJ...659..561M,2009MNRAS.396.2038B,2011MNRAS.413.2031M} and (ii) NS-NS mergers, can themselves lead to the creation of millisecond protomagnetars, other cosmic events, such as (iii) the accretion-induced collapse of white dwarfs \cite{1991ApJ...367L..19N,2006ApJ...644.1063D,2008MNRAS.385.1455M,2009arXiv0908.1127M}, or the merger of two white dwarfs \cite{2001MNRAS.320L..45K,2008MNRAS.385.1455M}, are also plausible progenitors.

Millisecond protomagnetars may emit a sizable fraction of their rotational energy via GWs \cite{2002PhRvD..66h4025C,2005ApJ...634L.165S,2009MNRAS.398.1869D,2011PhRvD..84b3002K}, although they may also lose angular momentum via magnetic processes (e.g., \cite{2006PhRvL..96c1101D,2006PhRvD..74j4026S}). Fast rotation can result in the emergence of rotational instabilities, especially if the protoneutron star is differentially rotating (see also Section \ref{section:protoneutronstar}). If a millisecond magnetar indeed drives a GRB, it needs to maintain fast rotation for durations comparable to the duration of the GRB, potentially allowing sufficient time for rotational instabilities to emerge and result in GW emission. If a significant fraction of the rotational energy ($\sim10^{-2}$M$_\odot\,$c$^2$; e.g., \cite{2001ApJ...550..426L}) is converted into gravitational radiation, the resulting GW signal could be detected to a distance of $\sim 100\,$Mpc with advanced detectors (see also Section \ref{section:protoneutronstar}).

The rotation rate of millisecond magnetars, and therefore the emitted GW strength and frequency, likely depends on the cosmic event they originate from. Accretion-induced collapse of white dwarfs that accrete mass and angular momentum from a non-degenerate companion may produce the most rapidly rotating protomagnetars. Due to continued accretion, such white dwarfs will be rapidly rotating upon collapse, resulting in a rapidly rotating protomagnetar, and a strong GW signal. On the other hand, the remnant of a binary white dwarf merger will likely collapse to a magnetar with a significant delay ($\sim10^6\,$yr), allowing the remnant to lose most of its angular momentum via stellar winds \cite{2007MNRAS.380..933Y,2012arXiv1207.0512S,2012ApJ...748...35S}. Binary white dwarf mergers are therefore not likely to be significant sources of GWs.

\section{GRB astrophysics with gravitational waves}
\label{section:astrophysics}

This section addresses the question: What can the detection of GW signals add to our understanding of the physics of GRBs and their progenitors? Detectable electromagnetic radiation, our main source of information on GRBs, is observable only from relatively large distances ($\gtrsim10^{12}-10^{13}\,$cm) from the central engine. On the other hand, GWs are created right at the central engine, and can convey information about it without being distorted or absorbed by matter on their way to the observer. Here we outline some of the questions of interest that could be addressed through observing the GW signature of GRBs.

\subsection{Compact binary coalescence}
\label{section:cbcphysics}

\begin{enumerate}[$\bullet$]
\item{\bf Progenitors of short GRBs} --
As compact binary mergers will be detectable from hundreds of Mpc with the Advanced LIGO-Virgo network \cite{2010CQGra..27q3001A}, within these distances it will be possible to confirm or rule out compact binaries as GRB progenitors. While identification based on electromagnetic signals is not always straightforward \cite{2012arXiv1204.4919Z}, the detection of GWs will be likely essential to unambiguously reveal the nature of the progenitor \cite{Cutler:2001}. By connecting each measured source with the presence or absence of detectable GW emission, and by exploring the connections between observed electromagnetic counterparts and their binary progenitors, the central engine will be probed in unprecedented ways. For example, observing GRBs in coincidence with those binaries for which numerical simulations predict the formation of accretion disks would indicate a connection between accretion disks and GRBs. Already, analyses of GW data from initial detectors targeting gamma-ray bursts have helped rule out binary mergers as the central engines of selected events \cite{070201,051103}.
\item{\bf Population prospects} --
The populations of short GRBs \cite{2007ApJ...664.1000B}, as well as compact binaries \cite{2010CQGra..27q3001A}, are highly uncertain. The observation of GWs from compact binary coalescences could be an effective way of determining the source population. Due to the very large field-of-view of GW detectors and the very weak beaming of GW emission, they can be used efficiently to locate practically all sources within their well-determined horizon distance of hundreds of megaparsecs. As source distances can also be determined using GWs (see next subsection), the binary coalescence population could also be mapped as a function of redshift, although only with the sensitivity of third-generation GW detectors \cite{2009arXiv0906.4151S}.
\item{\bf Cosmological parameters} --
The detection of a binary coalescence GW signal from an observed short GRB could be used to determine the distance and redshift of the source \cite{1986Natur.323..310S,1993PhRvD..48.4738M,2010ApJ...725..496N}. In fact, the determination can be done independently of the cosmological distance ladder. Distances of short GRBs can be reconstructed to a $\sim10-30\%$ precision for $\lesssim500\,$Mpc for NS-NS, and for $\lesssim1.5\,$Gpc for NS-BH binaries using the Advanced LIGO-Virgo detector network \cite{2010ApJ...725..496N}. Reconstructed source distances can then be used to accurately reconstruct the luminosity of short GRBs, as well as to test Hubble relation \cite{1986Natur.323..310S}, with one year of observation with advanced detectors potentially allowing for a $\sim2\%$ precision \cite{2006PhRvD..74f3006D}. Further, in combination with measurements of the cosmic microwave background, reconstructed binary merger distances could be used to constrain the dark energy equation of state \cite{2006PhRvD..74f3006D}.
\item{\bf Jet angular structure} --
The angular structure of relativistic jets from short GRBs is poorly constrained \cite{2011ApJ...732L...6R}. The rate of binary mergers detected through GWs, together with the rate of short GRBs for which a binary progenitor can be confirmed with GWs, could be used to determine the opening angle of short GRBs.

Further, the polarization of the inspiral GW signals in principle could be used to characterize the viewing angle of observed GRBs \cite{1538-4357-585-2-L89}. The GW polarization from a binary inspiral depends on the viewing angle compared to the rotational axis of the binary.
Towards the rotational axis, the GW signal is circularly polarized, while the polarization becomes elliptical for off-axis observers, eventually becoming linear for observers in the equatorial plane. For sufficiently strong GW signals from the inspiral phase, reconstructing the eccentricity of the GW polarization could provide information on the opening angle. With a large number of measurements, the angular structure of the jet could be mapped as well \cite{1538-4357-585-2-L89}. These correlations can be compared to similar constraints that might follow from the detailed multiband light curve of the blastwave \cite{2011ApJ...733L..37V}. Nevertheless, due to the relatively weak change in GW polarization with angle, the GW polarization may only be identified for large ($\gg1^\circ$) angular differences, and/or for very high SNR ($\gtrsim100$), making the utility of GW polarization limited.
\item{\bf Neutron Star Equation of State} --
Nuclear forces have a profound influence on the structure of NSs (e.g., \cite{1998nucl.th...4027A,2001ApJ...550..426L}). They determine, among other things, the relation between NS mass and radius, or the NS mass limit (for the rotating and non-rotating cases). The EOS of matter is poorly constrained at NS densities \cite{2008PhRvD..77b1502F}. The observation of NSs with masses up to $\sim2\,$M$_\odot$ \cite{2005ApJ...634.1242N} imply a stiff EOS (a NS with soft EOS would not be able to support this much mass).

The evolution of NS-NS/BH-NS mergers strongly depends on the NSs' nuclear EOS. Therefore, the GW signal from binary coalescences can be used to determine/constrain the EOS. For instance, if a NS is tidally disrupted during a binary merger, the orbital frequency at which tidal disruption occurs can be used to determine the radius of the NS. This, together with the NS mass reconstructed from the inspiral gravitational waveform \cite{1994PhRvD..49.2658C}, can be used to constrain the nuclear EOS \cite{2000PhRvL..84.3519V,2007PhR...442..109L,2010PhRvD..82d4049K,2011PhRvD..84f4018K,2012PhRvD..85d4061L,2012PhRvL.108i1101M}.

The tidal deformation of NSs in a binary system can affect the gravitational waveform even prior to the merger phase. The GW energy spectrum during the last few orbits of a NS-NS binary prior to merger can be used to determine the compactness ratio (mass/radius) of the NSs \cite{2002PhRvL..89w1102F,2012PhRvL.108i1101M}. For these last few orbits of the binary, the GW frequency is at a high but still reasonably sensitive frequency for LIGO-type detectors, making the analysis of these last few orbits feasible.

The tidal deformation of a NS in the inspiral phase can be described by one parameter (the so-called \emph{Love number} \cite{2008PhRvD..77b1502F}), which is effectively the ratio of the star's induced quadrupole moment to the quadrupole moment of the perturbing tidal gravitational field of the binary companion. Flanagan \& Hinderer \cite{2008PhRvD..77b1502F} showed that the nuclear EOS of NSs can be constrained even through the early inspiral phase due to the effect of tidal deformation on the waveform. Read et al. \cite{2009PhRvD..79l4033R} simulated the inspiral phase of NS-NS binaries, showing that the NS EOS can be constrained (the NS radius can be determined to within $\sim1$\,km precision, which, together with the NS mass, would rule out a part of the EOS parameter space) with advanced GW detectors for a source at a distance of $100\,$Mpc. BH-NS mergers can similarly be used to gain information on the NS EOS \cite{2011PhRvD..84j4017P,2012PhRvD..85d4061L}.

For some NS-NS mergers, collapse to a BH is delayed and a hypermassive NS is formed (see Section \ref{subsubsection:mergerphase}). During the merger, non-axisymmetric oscillation modes of the hypermassive NS are excited, resulting in GW emission \cite{1994PhRvD..50.6247Z,1999PhRvD..60j4021A,2002PhRvD..65j3005O,2002PThPh.107..265S,2005PhRvD..71h4021S,2011MNRAS.418..427S,2012PhRvL.108a1101B,2012PhRvD..86f3001B}. The waveform of the resulting gravitational radiation depends on the nuclear EOS. Consequently, the observed waveform can be used to significantly constrain the EOS \cite{2002PhRvD..65j3005O,2002PThPh.107..265S,2005PhRvD..71h4021S,2011MNRAS.418..427S,2012PhRvL.108a1101B,2012PhRvD..86f3001B}. The GW signature of hypermassive-NS oscillations could be detected with SNR$=2$ out to $20-45\,$Mpc (depending on the EOS), which may be sufficient to constrain the EOS (at such distances, the source can be confirmed with high accuracy via GWs from the inspiral phase) \cite{2012PhRvL.108a1101B}.

Further, it is possible that the inner part of a NS becomes strange quark matter at high densities. A star made of strange quark matter can be self-bound, a marked difference from hadronic NSs that are gravitationally bound \cite{Lattimer:2006}. The GW emission of a strange quark star is substantially different from that of hadronic NSs \cite{2007PhR...442..109L}, hence GWs could be used to determine whether NS interiors may contain strange quark matter \cite{2002MNRAS.337.1224A,2004MNRAS.349.1469O,2005PhRvD..71f4012L,2011PhRvL.107u1101S}.
\item{\bf Magnetic fields in protoneutron stars} --
Magnetic fields are suspected to play a significant role in the origin of short GRBs \cite{2008PhRvD..78b4012L,2010CQGra..27k4002D,2011ApJ...732L...6R,2012ApJ...755...80P}, and are known to vary between individual pulsars (e.g., from their spindown rate \cite{1977ApJ...215..302F}).
In the inspiral phase, internal magnetic fields in NSs (for realistic field strengths) have negligible effect on the dynamics of the binary coalescence \cite{2000ApJ...537..327I,2008PhRvD..78b4012L,2008PhRvD..78h4033B,2009MNRAS.399L.164G,2010PhRvL.105k1101C}. 
While magnetic tension in the NS can reduce tidal deformation, this effect will be too weak to be detectable with planned GW detectors for realistic pre-merger magnetic strengths ($\lesssim10^{14}\,$G) \cite{2009MNRAS.399L.164G}. For BH-NS binaries, magnetic fields seem not to alter the dynamics and therefore the gravitational waveform \cite{2010PhRvL.105k1101C}.

Magnetic fields can significantly affect the post-merger behavior of NS-NS binaries, if a hypermassive NS is formed after the merger. Upon merger, magnetic fields will be amplified via Kelvin-Helmholtz instabilities in the shear layer between the merging NSs \cite{2006Sci...312..719P,2008PhRvD..78h4033B}, or later via differential rotation \cite{2006PhRvL..96c1101D,2008PhRvD..78b4012L}. Even if magnetic fields are small prior to merger, the amplified fields can be strong enough to affect the evolution of the merger remnant and therefore GW emission. Magnetic fields in the hypermassive NS compete with GWs in dissipating angular momentum from the hypermassive NS. The decrease of angular momentum eventually results in the ``delayed collapse" ($\sim10\,$ms$-1\,$s, depending on the mass and the EOS) of the hypermassive NS into a BH (e.g., \cite{2008PhRvD..78b4012L,2012PhRvL.108a1101B}).

By probing the magnetic properties of a large number of NSs in binary mergers, one could determine, e.g., the highest achievable magnetic field frozen in a stationary NS (\cite{Lattimer:2006}; e.g., via the time difference between binary merger and GRB).
\item{\bf Accretion disks} --
The large-amplitude, quasiperiodic GW emission from instabilities in the accretion disk, probably together with the detection of the earlier inspiral phase from the binary, can be utilized to reconstruct properties of the accretion disk, even if no electromagnetic signal is observed from the source. Beyond reconstructing the mass and lifetime of the accretion disk, the development of the instabilities is sensitive to the properties of the accretion disk (see Section \ref{section:accretiondiskinstabilities}), i.e. the development of the instabilities by itself can already be informative.
\end{enumerate}

\subsection{Core collapse}
\label{section:collapsarphysics}

\begin{enumerate}[$\bullet$]
\item{\bf Progenitors of long GRBs} --
It is believed that the origin of at least some long GRBs are the core collapses of massive stars \cite{2006ARA&A..44..507W}. This could be confirmed if the GWs in coincidence with a long GRB progenitor were detected. Nevertheless, if no GW emission is detected in coincidence with long GRBs, this does not rule out the core-collapse model, although may constrain core-collapse dynamics. If GWs are detected in coincidence with a long GRB, the internal evolution of a core collapse, which may largely depend on the progenitor, could also be examined. For example, the formation of a protomagnetar or a BH-torus system, or the fragmentation of the massive stellar core may be differentiated via GWs.
\item{\bf Neutron star equation of state} --
The mass, radius and rotation rate of NSs have a profound effect on their potential GW emission through rotational instabilities. Detecting the GW signature of these instabilities can be used to constrain the mass-radius relation, rotation, and therefore the EOS, of NSs (e.g., \cite{2011PhRvD..83f4031G}).

Further, rotational instabilities in NSs result in qualitatively different GW emission for conventional and strange quark NSs \cite{2011GReGr..43..409A,2005PhRvL..95u1101O}. Whether a NS is composed of strange quark matter could be inferred from the detected GW signature of the NS's rotational instability.
\item{\bf Neutron star internal physics} --
The evolution of the gravitational waveform from protoneutron star rotational instabilities strongly depends on the nuclear EOS at NS densities, as well as the physical parameters of the protoneutron star. Differential rotation inside a protoneutron star \cite{2005PhRvD..71b4014S,2007PhRvL..98z1101O,2007CQGra..24..139O,2008A&A...490..231S}, temperature (e.g., \cite{2009PhRvD..79j4003B}), viscosity and neutrino cooling \cite{2009PhRvD..79j4003B} may all leave their mark on the evolution of rotational instabilities and the resulting GW emission. Further, if a hypermassive (or supramassive; see Section \ref{subsubsection:mergerphase} for their definition) NS collapses into a BH due to either losing angular momentum or accretion, the rotational frequency prior to collapse provides information on the NS EOS \cite{2007PhR...442..109L}. This rotational frequency may be inferred from the GW signal from the rotating NS \cite{2011PhRvD..83f4031G,2012arXiv1207.3805P}.
\item{\bf Accretion physics} --
Because GW emission from massive stellar collapses can be produced by the accretion onto the central object following the collapse, the GW signal therefore can provide insight into accretion physics: (i) the mechanisms that transport angular momentum; (ii) instabilities; and (iii) interaction with the black hole.

For BH-torus systems the strength of viscosity (and possibly the strength of other processes through which the torus loses angular momentum) can be determined, as these are competing effects for angular momentum loss: The total loss of angular momentum can be inferred from the rate of change of the GW frequency, while the loss through GWs can be inferred from the GW signal amplitude.

Accretion of matter by the protoneutron star results in extended GW emission due to the intake of angular momentum \cite{1998ApJ...501L..89B,2007PhRvD..76h2001A}. As the time scale and nature of accretion is likely different for accretions from supernova fallback material or material from a companion star, the extended GW signal from rotational instabilities will carry important information on the accretion mechanism.
\item{\bf Magnetic fields in neutron stars} --
Sufficiently strong magnetic fields inside a protoneutron star can be competing with GWs in radiating angular momentum away. As NS spindown may be reflected in the GW frequency, measuring the GW amplitude and NS spindown may provide information on the strength and nature of magnetic fields present in the NS.
\end{enumerate}

\subsection{Magnetars}
\label{section:magnetarphysics}

The detection of GWs in coincidence with a giant flare from a magnetar could provide information on the processes that lead to flaring in magnetars (i.e. tectonic activity \cite{1996Natur.382..518C} or global reconfiguration \cite{2001MNRAS.327..639I} vs origin from the magnetosphere \cite{2006MNRAS.368L..35L}). Non-detection, nevertheless, does not necessarily rule out these models. Upon detection, the frequency of GWs from NSs excited by, e.g., starquakes, could be used to infer the NS mass, radius and EOS \cite{2011MNRAS.414.3014C,2012MNRAS.423..811C,2012PhRvL.108t1101S}, or potentially even the strength of magnetic fields inside the NS \cite{2011PhRvD..83h1302K}.

\subsection{Millisecond protomagnetars}
\label{section:protomagnetarphysics}

The detection of millisecond magnetars' GW signature in coincidence with a GRB would confirm their presence and role as the central engine behind some GRBs. Further, the GW signature of millisecond magnetars may indicate their origin (e.g., typical rotation rates may differ for different mechanisms; see Section \ref{section:millisecondmagnetars}). Advanced GW detectors will also be able to determine, out to $\sim450\,$Mpc, whether the progenitor of a short GRB was a binary merger of a NS with another NS or a BH \cite{2010CQGra..27q3001A}. The confirmation of a compact binary merger would rule out the other scenarios in which millisecond magnetars can form, e.g., white dwarf accretion-induced collapses, or white dwarf binary mergers.

\section{Observational strategies and prospects}
\label{section:observationalstrategies}

The success of detection of GWs from GRB progenitors depends on the strength of the emitted GWs, the sensitivity of GW observatories, as well as the observation strategies used to separate GW signals from the background. The search strategy, e.g., the use of multiple messengers, can also add to the information one can extract from the source.

Various transient GW search strategies exist, aiming to find the GW signature of GRBs. These strategies differ mainly in their prior assumptions on the gravitational waveform. In the most general case, one can look for so-called GW bursts with minimal assumptions on the waveform, namely defining a maximum signal duration and bandwidth \cite{Klimenko:2005,Xpipeline1367-2630-12-5-053034,RollinsThesis}. These generic burst searches can be modified to include additional information about the source. For instance if one considers an accretion-type emission, circular (on-axis) or linear (off-axis) polarizations can be required from GW signals, increasing detection sensitivity for signals satisfying these conditions. Other model-dependent assumptions include, e.g., GW signals rapidly evolving characteristic frequency \cite{2010CQGra..27s4017C}, such as in the case of compact binary mergers.

For compact binary coalescences, the gravitational waveform for a given mass configuration can be accurately calculated for the initial inspiral and final ringdown phases, and to some extent the merger phase \cite{2011PhRvD..83l2005A}.
For such systems with properties in a highly limited parameter space, it is beneficial to use a signal waveform template bank and perform matched filtering based analysis (e.g., \cite{2010PhRvD..82j2001A}).

Below we review the GW search strategies targeting the progenitors of GRBs. We indicate the prospects of such searches through presenting projected sensitivity estimates based on observational results with initial detectors. Advanced detectors will be $\sim10\times$ more sensitive than initial detectors, the exclusion distances therefore will be $\sim10\times$ greater, corresponding to a $\sim$1000-fold improvement in achievable source rate constraints. Further, advanced detectors will be sensitive in a wider frequency band. Additionally, the advanced detector network will eventually be larger, with the inclusion of KAGRA and probably LIGO India. These improvements will further increase the achievable sensitivity of the advanced detector network \cite{2012arXiv1210.6362N}, and will further improve signal parameter reconstruction (e.g., source direction; \cite{2011PhRvD..83j2001K}) \cite{2012arXiv1210.6362N}.

\subsection{Gravitational wave search strategies}

Search strategies for GW transients can be divided into four main categories based on their use of other messengers.
\begin{enumerate}[1]
\item \textbf{Untriggered} searches, in which no information is used from other messengers. The analyses use GW data alone to find astrophysical sources (e.g., \cite{2010PhRvD..81j2001A,2010PhRvD..82j2001A}).
\item \textbf{Externally-triggered} searches, in which one specifically looks for GW signals from sources confirmed via other messengers. For instance one can search for GW signals from the progenitors of detected GRBs, using their location, time, and other parameters (e.g., \cite{2005PhRvD..72d2002A,S2S3S4,2008ApJ...681.1419A,2010ApJ...715.1438A,2010ApJ...715.1453A,2011ApJ...734L..35A}).
\item \textbf{Electromagnetic follow-up} searches, in which GW signal candidates are used to trigger electromagnetic follow-up searches with other telescopes (e.g., \cite{2003ApJ...591.1152S,2007AAS...211.9903P,2008CQGra..25r4034K,2010AAS...21540606S}).
\item \textbf{Multimessenger} searches, in which one uses (typically sub-threshold) signal candidates from multiple observatories of different messengers in a joint search (e.g., \cite{2009IJMPD..18.1655V,Baret20111,PhysRevLett.107.251101,2012PhRvD..85j3004B}).
\end{enumerate}
In the following we briefly review some of the GW search strategies and past GW searches for GRBs, organized along the above categories.

\subsubsection{Untriggered searches}
\label{section:allsky}

--- Untriggered transient searches provide a generic way to identify plausible GW sources without relying on the detection of other messengers. While GRBs are highly beamed (e.g., \cite{2001ApJ...562L..55F,2005Natur.437..845F,2006ApJ...650..261S,2006ApJ...653..468B,2006ApJ...653..462G,Liang2007,2012ApJ...756..189F}), the beaming of GW emission from GRB progenitors is weak (e.g., \cite{1538-4357-585-2-L89}). For realistic GRB beaming angles, there will likely be more observations of GWs from off-axis GRB progenitors than GW observations triggered by GRBs \cite{2012arXiv1206.0703C}.

To help quantify the source population detectable with untriggered (and other) searches, it is useful to characterize the sensitivity of a GW observatory to a GW source with the so-called \emph{horizon distance} $D_h$. The horizon distance is defined as the distance at which a GW source from optimal orientation (such that emission towards the Earth is the strongest) and optimal location (i.e. the most sensitive direction of the GW detector) is observed with a given SNR (typically chosen to be $\rho=8$ for untriggered searches; see also Section \ref{subsubsection:extrig}) by a single GW detector. Compared to this optimal orientation and location, the orientation and direction-averaged distance at which the source produces the $\rho=8$ is $\sim D_h/2.26$ (both direction and orientation-averaging contribute a factor $\sim1.5$).

For a NS-NS binary with 1.35~M$_\odot$ mass for each NS, the projected horizon distance for untriggered searches with advanced detectors (upon non-detection) is \cite{2010ApJ...716..615O}
\begin{equation}
D_h^{untrig.}\simeq 450\,\mbox{Mpc}\,(\mathcal{M}_c/1.2\,\mbox{M}_\odot)^{5/6},
\label{eq:Dh}
\end{equation}
where $\mathcal{M}_c=(m_1m_2)^{3/5}/(m_1+m_2)^{1/5}$ is the so-called chirp mass (i.e. the horizon distance only depends on this specific combination of the masses of the two compact objects). For higher-mass binaries, the horizon distance is somewhat lower than the prediction of Eq. \ref{eq:Dh} (see \cite{2012arXiv1206.0703C}).

\subsubsection{Externally triggered searches}
\label{subsubsection:extrig}

--- External triggers in GW searches provide additional information and increased search sensitivity for GW analyses \cite{1993ApJ...417L..17K,FMR99}. For example an external GRB trigger reduces the temporal and spatial extent in which one has to search for a GW signal \cite{2009RPPh...72g6901A}. A detected external trigger can also be used to set constraints on the GW signal. For example a short GRB implies that the GW signal is probably a binary merger.

Chen \& Holz \cite{2012arXiv1206.0703C} estimates that for GRB beaming factor $f_b\leq7.5$ (corresponding to opening angle $\theta=30^\circ$), the number of triggered and untriggered detections of GWs from GRB progenitors would be comparable. Further, even for greater beaming factors (i.e. smaller opening angles), externally triggered searches will enable the detection of some additional, particularly interesting, sources. The extra information in external triggers can be taken into account in the horizon distance by effectively lowering the SNR threshold (i.e. by keeping the false-alarm rate constant). The results of Chen \& Holz imply that the sensitivity of externally triggered searches is greater than that of untriggered searches by a factor $\sim1.3$ (for inspiral searches; this factor is similar for other types of searches as well), i.e.
\begin{equation}
D_h^{ext.trig.}\approx 1.3D_h^{untrig.}.
\end{equation}
The same sensitivity increase was found by Dietz et al. \cite{2012arXiv1210.3095D}. While this increased horizon distance means that many more sources fall within observable reach, the actual number of detected sources will be decreased by the beaming of the electromagnetic (or other) signal, as well as the sky coverage of the available electromagnetic telescopes. On the other hand, all triggered sources will be detected face-on, while untriggered sources can take any orientation, which adds to the sensitivity of triggered searches compared to untriggered searches (for binary mergers, the GW amplitude is $\sim1.5$ times greater face-on than averaged over all directions). Chen \& Holz \cite{2012arXiv1206.0703C} find that the externally-triggered observation rate is a sizable addition, although less than, the untriggered GW observation rate, while Kelley et al. \cite{2012arXiv1209.3027K} find that GRB-triggered searches present no appreciable addition to the total rate \cite{2012arXiv1209.3027K}.
While both of these studies approximate GW background as purely Gaussian, real GW background features a non-Gaussian tail that may increase the importance of externally triggered and follow-up searches.

Previous GRB-triggered GW searches include searches for both unmodeled GW bursts and compact binary mergers. For example the short GRBs 070201 and 051103 had electromagnetic positions overlapping nearby galaxies (Andromeda and M81, respectively). No binary merger counterpart was found in either case, ruling out a binary progenitor in those galaxies, favoring an SGR model (a binary progenitor from a more distant galaxy is not ruled out, albeit it is unlikely) \cite{0004-637X-681-2-1419,2012ApJ...755....2A}. Nevertheless, it is worth noting that, given their expected rate of occurrence, a binary merger within the distance of Andromeda or M81 ($<4\,$Mpc) would be extremely unlikely. For GRB-triggered GW searches with initial LIGO/Virgo, see also, e.g.,  \cite{2012ApJ...760...12A}.

Several GW searches aimed to identify GWs in coincidence with SGR flares  \cite{2007PhRvD..76f2003A,2008PhRvL.101u1102A,2009ApJ...701L..68A,2011ApJ...734L..35A}. Most recently Abadie \emph{et al.} \cite{2011ApJ...734L..35A} used 1279 flares from six magnetars as triggers, aiming to identify GW signals from neutron-star $f$-mode ringdowns or other GW-producing mechanisms. The search set constraints on the energy of GW emission from these flares comparable to some giant flares' electromagnetic energies, and an order of magnitude below previously existing limits. For a nearby magnetar (SGR 0501+4516) located at $\sim1-2\,$kpc that emitted a large number of flares during the observation period, the obtained upper limit on $f$-mode GW emission was $\sim 10^{47}$~erg. The upper limit for GW emission at $\sim10^2$~Hz, i.e. within most sensitive frequency band of LIGO/Virgo, was $\sim3\times10^{44}$~erg, a promising limit compared to some theoretical upper limits on GW emission from SGRs \cite{2011PhRvD..83j4014C}. With Advanced LIGO/Virgo, one can expect, beyond more stringent limits on $f$-mode emission, constraints on the maximum emissible \emph{total} energy from other NS oscillation modes (we note that these upper limit are constraining only if a substantial fraction of total energy goes into GWs, which is debated; e.g., \cite{2012arXiv1206.6604C}).

\subsubsection{Electromagnetic follow-up}

--- With the rise of a global GW detector network, it became possible to reconstruct the direction of a GW signal candidate, albeit with a relatively large uncertainty. Such reconstruction enables the use of GW signal candidates in triggering electromagnetic follow-up observations \cite{2003ApJ...591.1152S,2007AAS...211.9903P,2008CQGra..25r4034K}. There are various scientific advantages of such a follow-up search:
\begin{enumerate}[$\bullet$]
\item As electromagnetic observations cannot continuously cover the whole sky with high sensitivity, GWs can be used to guide these telescopes and point towards the right direction at the right time.
\item Often the most interesting and strongest emission from an electromagnetic transient occurs in a short period after its onset. Since GW emission is, in many cases, expected to precede the onset of electromagnetic emission, telescopes have the chance to commence follow-up observation in a very early emission stage, or even catch the onset of the electromagnetic event.
\item GW signal candidates will often have too low significance to be unambiguously identified as extraterrestrial signals. The observation of an electromagnetic follow-up event can enhance the significance of the joint observation, making detection more probable.
\item Similarly to other multimessenger searches, information from the different messengers can enhance the information (and therefore science) that one can extract from the source.
\end{enumerate}
Electromagnetic follow-up searches of GW event candidates is one of the most promising and one of the fastest-evolving areas of GW astrophysics (see \cite{2012A&A...539A.124L,2012A&A...541A.155A,2012arXiv1205.1124E} for recent observational development with the LIGO-Virgo GW detectors). See Section \ref{section:EMcounterpart} below for a more detailed description of the potentially detectable electromagnetic counterparts and search strategy.

The sensitivity of electromagnetic follow-up searches will differ from the sensitivity of externally triggered searches due to (i) the limited reach of electromagnetic transient observations and (ii) the finite sky area that is scanned by observatories following up the GW event candidate. Taking (i) into account in defining the horizon distance, one gets
\begin{equation}
D_h^{follow\mbox{-}up}\approx \mbox{min}\left\{D_h^{ext.trig.},D_{\mathrm{EM}}\right\},
\end{equation}
where $D_{\mathrm{EM}}$ is the reach of the electromagnetic follow-up observatory. Note that $D_h^{ext.trig.}$ refers to on-axis sources, while follow-up observations are the most interesting for off-axis sources (due to the greater source rate). For off-axis directions, GW emission is typically weaker by a factor $\sim1.5$. The rate of follow-up observations further depends on sky coverage of the follow-up observatories, as well as the beaming of the electromagnetic signal. For a survey of the electromagnetic detectability of GW event candidates, see \cite{2012arXiv1210.6362N}.

\subsubsection{Multimessenger analyses}

--- Multimessenger analyses generalize the idea behind externally triggered searches by combining information from sub-threshold signal candidates of various messengers into one, more powerful analysis. While in many cases each type of sub-threshold messenger would individually have too low significance to be identified as an astrophysical signal of interest, the combination of individual significances can greatly enhance the analysis' potential of identifying astrophysical events (see Section \ref{chapter:GWHENsearch} for a multimessenger search).

The \emph{Astrophysical Multimessenger Observatory Network} (AMON) initiative \cite{AMON} plans to combine sub-threshold triggers of multiple messengers in a low-latency coherent multi-messenger analysis. AMON is planned to combine multiple messengers to identify candidates that would be sub-threshold events for individual observatories. The extracted information would be used, besides claiming detection itself, to initiate electromagnetic follow-up observations, which can add to the number of coincidentally observable messengers.

The sensitivity of multimessenger searches depends on the false alarm statistics of the different detectors included in the search, as well as the types of messengers. Transient messengers with amplitude signatures, such as GWs or high-statistics photon signals, will have a relatively well defined horizon distance, beyond which the detection of such a messenger is highly unlikely. Other, discrete messengers, such as high-energy neutrinos, will have no such horizon distance, since there is a finite probability of detecting at least one neutrino even from a very distant source with expected signal strength $\ll1$ neutrino. These distant sources can significantly contribute to the total number of detectable multimessenger sources, especially in the case in which their presence can be confirmed with other messengers (see, e.g., \cite{2012PhRvD..85j3004B}). For such discrete messengers, the GW-multimessenger horizon distance is comparable to that of externally triggered searches:
\begin{equation}
D_h^{multimessenger}\sim D_h^{ext.trig.}.
\end{equation}
The rate of detected multimessenger sources will be limited by beaming, and the probabilistic nature of observing at least 1 discrete messenger (e.g., high-energy neutrino  \cite{PhysRevLett.107.251101}).

\subsection{Electromagnetic counterparts}
\label{section:EMcounterpart}

This section outlines the observable electromagnetic emission of GRB progenitors, focusing on counterparts other than gamma rays, and their utility in GW searches.

Due to the highly beamed emission of most GRBs\footnote{Beaming of short GRBs in general is not well constrained \cite{2007PhR...442..166N}. Nevertheless, beaming has been confirmed for a few short GRBs \cite{2006ApJ...653..468B,2012ApJ...756..189F}, indicating a highly beamed jet with half opening angles $\lesssim 10^\circ$.}, the majority of GRBs are off-axis, i.e. their prompt emission cannot be observed from the earth. Nevertheless, weaker, off-axis electromagnetic emission from GRB progenitors could be detected during GW follow-up observations, if the electromagnetic emission is present beyond a few minutes after the prompt emission (i.e. the delay between detecting a GW signal candidate and pointing a telescope towards the expected source location).

Most compact object mergers observed by LIGO/Virgo will not be accompanied by observable GRBs due to the small gamma-ray beaming angle.  For this reason, the last few years have seen extensive investigation into rapid follow-up search strategies for more isotropic, but more difficult to identify, electromagnetic counterparts \cite{2007AAS...211.9903P,2008CQGra..25r4034K,2012ApJ...748..136C,2012A&A...539A.124L,2012A&A...541A.155A,2012ApJ...759...22K}. Follow-up searches aim to minimize the time between the identification of a GW event candidate and the start time of the observation with electromagnetic telescopes in the required directions. The time delay of sending a request for follow-up search to electromagnetic telescopes after a GW event was $\sim$tens of minutes for initial searches (mostly due to the manual confirmation of event selection). This delay will likely be reduced to $\sim$minutes for searches with advanced detectors \cite{2012SPIE.8448E..0QS}.

The uncertainty in direction reconstruction of GW event candidates with the planned advanced detector network will likely be in the $\sim$tens of square degrees range (e.g., \cite{2011PhRvD..83j2001K,2012arXiv1210.6362N}). With such an uncertainty, telescopes with large fields of view will need to be utilized, focusing on electromagnetic emission that is ongoing for hours-days so the telescopes can scan through the interesting sky area. Existing galaxy catalogs can also be used to largely decrease the sky area that needs to be mapped (see Section \ref{subsection:hostgalaxies}). We note here that, besides GW observations, the observation of the prompt gamma-ray emission itself has a position resolution of a few degrees, i.e. comparable to the resolution of GW observations. Afterglow observations are therefore also important for the direction reconstruction of observed GRBs.

Beyond direction reconstruction, GRB afterglows also promise to carry important information connected to GW radiation. Corsi \& M\'esz\'aros \cite{2009ApJ...702.1171C} pointed out that the observed X-ray plateau in the afterglow of some GRBs may be connected to the formation of a highly magnetized millisecond pulsar that can lose angular momentum through GWs during a 10$^3-10^4\,$s period (see also \cite{2001ApJ...552L..35Z}).

Several promising emission models have been identified which predict longer-lived, weakly (or not at all) beamed emission that can be detected on a distance scale comparable to the sensitivity range of advanced GW detectors. In the following list of these models, we focus on optical counterparts of compact binary coalescences. For a survey of the detectability of some of these emission models, see, e.g., \cite{2012arXiv1210.6362N}.
\begin{enumerate}[$\bullet$]
\item {\bf Electromagnetic remnants from compact binary mergers} --- NS-NS and NS-BH mergers can eject energetic (sub-relativistic or relativistic) outflows. These energetic outflows can interact with matter in the interstellar medium, creating a long-lasting radio signal \cite{2011Natur.478...82N,2012arXiv1204.6242P}. Radio remnants from dynamically ejected sub-relativistic material can appear months to years after the binary merger, and can last for several years \cite{2012arXiv1204.6242P}. A deep radio survey can identify these radio remnants out to $\sim300\,$Mpc \cite{2011Natur.478...82N,2012arXiv1204.6242P}. Mildly/ultra-relativistic outflows may create even brighter radio emission, on the time scale of weeks \cite{2012arXiv1204.6242P}. The detectability of the radio transient strongly depends on circum-merger density, and therefore the signal may be lower for binaries that left their galaxy prior to merger.

    An interesting case is shocks generated within the NSs upon the merger of a NS-NS binary \cite{2011PhRvL.107e1102S,2012PhRvD..86f4032P}, which can plausibly drive ultra-relativistic outflows. These shock waves, besides heating the NSs, result in shock breakout from the NSs' surface, driving a nearly omnidirectional, ultra-relativistic outflow \cite{2012arXiv1209.5747K}. The outflow decelerates in the ambient medium, resulting in a bright X-ray flare seconds after the merger that could be detectable with, e.g., Swift XRT \cite{2005SSRv..120..165B}. The flare gradually changes to optical and radio as the outflow decelerates, potentially detectable with, e.g., Pan-STARRS \cite{2002SPIE.4836..154K} and EVLA \cite{2011ApJ...739L...1P}.
    
\item {\bf Macronovae/kilonovae} --- Matter ejected from compact binary encounters may produce electromagnetic optical or near-infrared transients that are suitable for follow-up searches of GW signal candidates \cite{1998ApJ...507L..59L,2005astro.ph.10256K,2005ApJ...634.1202R,2010MNRAS.406.2650M,2011ApJ...734L..36S,2011ApJ...736L..21R,2011ApJ...738L..32G,2012ApJ...746...48M,2012arXiv1204.6240R,2012arXiv1204.6242P}. Matter ejected by binary mergers (tidal tails or accretion disk outflows; e.g., \cite{2005astro.ph.10256K}), initially at nuclear densities, expands and undergoes r-process nucleosynthesis, producing heavier, radioactive elements. The decay of such elements produces an isotropic emission that lasts for days, named ``macronova" \cite{2005astro.ph.10256K} or ``kilonova" \cite{2010MNRAS.406.2650M}.
    This is an appealing model, because it suggests that virtually all LIGO detected NS-NS or BH-NS mergers would have detectable optical counterparts. However, finding these counterparts in the large GW directional uncertainty region would likely require a combination of powerful survey telescopes and follow-up spectroscopy, as used by Palomar Transient Factory, and Pan-Starrs, or the future LSST \cite{2012IAUS..285..358M,2012arXiv1209.3027K,2012arXiv1204.6242P,2012ApJ...746...48M}.
\item {\bf X-ray/optical afterglow} --- The relativistic outflow from a GRB central engine interacts with matter in the circumburst medium, emitting a long-lasting electromagnetic afterglow in a wide range of frequencies (e.g., \cite{2011ApJ...733L..37V,2012ApJ...746...48M}). At X-ray and optical frequencies (lasting from minutes to a day), afterglow emission is the strongest shortly after the prompt GRB in the direction of the relativistic outflow (i.e. on-axis), and rapidly decays afterwards \cite{2011ApJ...733L..37V}. On-axis X-ray emission is detectable (e.g., with the Swift XRT telescope) over a period of $\sim1\,$day after the prompt burst for source distances out to the horizon distance of GW telescopes \cite{2012ApJ...759...22K}. Off-axis X-ray and optical emission, far from the axis of rotation, peaks at a significantly lower flux, making the detection of off-axis X-ray less useful in multimessenger searches \cite{2011ApJ...733L..37V}. For radio frequencies around $\sim10\,$GHz, on-axis emission for long GRBs peaks at 3-6 days in the source rest frame after the prompt burst \cite{2012ApJ...746..156C}, while off-axis emission peaks on a similar time scale.
    
\item {\bf Magnetic interaction between BH \& NS} --- For NS-BH binaries, once the BH enters the magnetosphere of the NS, it can interact with the NS magnetic field \cite{2011ApJ...742...90M}. For strong enough NS magnetic fields ($\sim10^{12}\,$G), such an interaction results in the copious emission of electromagnetic radiation, via a mechanism analogous to the Blandford-Znajek mechanism \cite{1977MNRAS.179..433B}. The electromagnetic radiation may resemble an extremely short ($\mathcal{O}(\mbox{ms})$) GRB , and could be detected from distances of $\gtrsim60\,$Mpc using the Swift or Fermi satellites \cite{2011ApJ...742...90M}. Although this mechanism produces a short electromagnetic emission, it is expected to be less beamed than the ``standard" gamma-ray emission.
\item {\bf Magnetic interaction between NS \& NS} --- The magnetic interaction of NSs in NS-NS binaries can drive an electromagnetic (X-rays or gamma-rays) signal within the last few seconds prior to merger through electric dissipation \cite{2012ApJ...755...80P}. As the non-magnetic NS moves through the magnetosphere of the other, highly magnetized NS ($\sim10^{12}-10^{14}\,$G), the induced electromagnetic force across the non-magnetic NS sets up a circuit connecting the two stars. Some of this electric potential dissipates either in the surface layer of the magnetic NS or in the space between the two stars, depending on the resistivity of the space between the two stars. A part of the dissipated energy is carried away through electromagnetic emission, which can reach up to $\sim10^{44}\,$erg$\,$s$^{-1}$ in the last $\lesssim1\,$s prior to merger \cite{2011PhRvD..83l4035L,2012ApJ...757L...3L}.
\item {\bf Neutron star flares induced by tidal crust cracking} --- NSs in inspiraling binary systems are subject to tidal deformation. When this deformation exceeds a critical level, the crust of the NS cracks, possibly resulting in a NS flare similar to observed flares \cite{1992ApJ...398..234K,2010PhRvD..82b4016P,2010ApJ...723.1711T,2012ApJ...749L..36P} (see Section \ref{subsection:magnetars} on magnetar giant flares). Such flare would represent an isotropic electromagnetic counterpart for binary mergers. Observed short-GRB precursors may be created by crust cracking \cite{2010ApJ...723.1711T}.
\end{enumerate}

\subsection{Neutrino counterpart}

Many GW sources, and in particular GRBs, are expected to be copious emitters of neutrinos \cite{0004-637X-657-1-383,0004-637X-690-2-1681,1997PhRvL..78.2292W,2001PhRvL..87q1102M,ET,P1000062}. Astrophysical neutrino emission is expected within two main sub-groups based on two distinct emission mechanisms, with two distinct energy ranges. MeV energy neutrinos (with energies $\epsilon_\nu \lesssim 100\,$MeV) are produced in the extremely hot, dense central regions of core-collapse supernovae and probably GRBs. High-energy neutrinos (with energies $\epsilon_\nu \gtrsim 100\,$GeV) are expected to be emitted by shock accelerated particles in relativistic outflows driven by the central engine of the GRB (e.g., \cite{1997PhRvL..78.2292W}). So far only MeV energy astrophysical neutrinos have been confirmed, and in one instance, for supernova 1987A \cite{PhysRevLett.58.1490,PhysRevLett.58.1494}.

One of the advantages of joint GW-neutrino searches \cite{PhysRevLett.107.251101,2012PhRvD..85j3004B} is that GW and neutrino detectors (MeV and high energy) continuously observe the whole sky (one high-energy neutrino detector observing half the sky), recording signal candidates without the need to ``point'' the detector in a particular direction. Such full sky coverage is of special importance for multimessenger searches as the sky coverage for each messenger needs to overlap for a joint search. Further, while electromagnetic follow-up searches require low-latency response from electromagnetic telescopes, GW-neutrino observations can be performed with high latency without loss of information. Nevertheless, fast analysis of joint GW-neutrino observations can also enable the electromagnetic follow-up of joint event candidates.

\subsubsection{MeV neutrinos}

--- In core-collapse supernovae, most of the released gravitational binding energy is emitted in a burst of $\sim$MeV neutrinos (e.g., \cite{2010A&A...517A..80F}). Both MeV neutrino \cite{PhysRevLett.104.251101,2011arXiv1108.0171I} and GW emissions \cite{Ott2008,0004-637X-707-2-1173} are expected to commence near core bounce within a few milliseconds ($\lesssim 10$~ms). This temporal correlation is orders of magnitude tighter than correlation with electromagnetic signals \cite{Baret20111,P1000062}, and can greatly enhance the sensitivity of a search for joint GW-MeV neutrino sources. MeV neutrino emission peaks within a fraction of a second after bounce, while emission continues for up to tens of seconds as the protoneutron star cools and contracts \cite{2011arXiv1108.0171I,2010A&A...517A..80F}.

The joint detection of GWs and MeV neutrinos could provide constraints on the core-collapse supernova mechanism as well as information on the properties of matter at high energies and densities \cite{P1000062}. For example the neutrino spectrum from core-collapse supernovae depend both on the nuclear EOS \cite{2009A&A...496..475M} as well as the spin of the core \cite{0004-637X-685-2-1069}. One can break this degeneracy using the additional information available in the GW channel, inferring information on both the EOS and the spin of the core. A potential secondary collapse due to hadron-quark phase transition in the protoneutron star during its post-bounce evolution \cite{PhysRevLett.102.081101} would also result in characteristic neutrino and GW emissions \cite{PhysRevLett.102.081101,MNR:MNR14056}.

Long and short GRB central engines are also probably strong emitters of MeV neutrinos \cite{0004-637X-657-1-383,0004-637X-690-2-1681,2011ApJ...737....6S,2012arXiv1206.5927S,2012arXiv1206.5927S}. As the progenitors of long GRBs are likely the progenitors of core-collapse supernovae as well \cite{hjorth:11,2006ARA&A..44..507W,modjaz:11}, emission from long GRBs may commence similarly to the case of core-collapse supernovae. In the core-collapse scenario, neutrino emission from the core ceases upon black hole formation, while a newly formed accretion disk will become a significant source, as a large fraction of the accretion energy can be carried away by neutrinos \cite{0004-637X-657-1-383}.

Compact binary mergers -- the likely progenitor of short GRBs -- are expected to emit MeV neutrinos \cite{0004-637X-657-1-383,0004-637X-690-2-1681} from the accretion disk around the central black hole. In the case of NS-NS binary mergers, a hypermassive NS may form prior to BH formation, and further produce neutrino \cite{0004-637X-690-2-1681} as well as GWs \cite{2008PhRvD..78h4033B} emission. MeV neutrino emission from compact binary mergers is expected to commence at the beginning with the merger phase, in coincidence with the GW burst due to the merger. This GW burst signal is preceded by the much longer inspiral phase, and followed by the ringdown phase. Such temporal coincidence can be utilized in a joint search. Nevertheless, for typical distances of binary mergers, MeV neutrinos would be difficult to detect \cite{2003MNRAS.342..673R}.

Several large-scale MeV neutrino detectors are currently in operation. These include Super-Kamiokande (Japan) \cite{2003NIMPA.501..418T}, KamLAND (Japan) \cite{KamLAND}, LVD (Italy) \cite{Aglietta:272638}, Borexino (Italy) \cite{Alimonti2002205} and Baksan (Russia) \cite{2009arXiv0910.0738N}. Further, the IceCube high-energy neutrino detector \cite{IceCubeAhrens2004507} is also capable of detecting bursts of MeV neutrinos \cite{2011arXiv1108.0171I}, albeit without the ability to reconstruct the source direction. Recognizing the importance of multimessenger observations, Super-Kamiokande, LVD, Borexino and IceCube are members of the Supernova Early Warning System (SNEWS) \cite{2004NJPh....6..114A}. These observatories send real-time triggers of detected supernova candidate events, which are distributed to the astronomer community to allow for low-latency follow-up electromagnetic (or other) searches.

The most sensitive current MeV neutrino detectors (Super-Kamiokande and IceCube) are able to detect the MeV neutrino signal of a supernova from up to $\sim$100~kpc \cite{2009PhRvD..80h7301H}. The expected event number within this distance, however, is relatively small outside of the Milky Way. Planned megaton detectors, such as LBNE \cite{2010JPhCS.203a2109M} or Hyper-Kamiokande \cite{2003nipb.conf..307N}, could detect supernovae from up to $\lesssim$10~Mpc \cite{2005PhRvL..95q1101A} with a supernova rate of $\sim1$ per year. Multimessenger searches of MeV neutrinos and GWs could further increase these detectable supernova rates \cite{2010CQGra..27h4019L}, and would provide increased confidence in a detected signal. This can be especially important when no electromagnetic counterpart is detected.

\subsubsection{High-energy neutrinos}
\label{chapter:GWHENsearch}

--- Non-thermal, high-energy ($\gg$GeV) neutrinos are thought to be created within relativistic outflows driven by the central engine \cite{1997PhRvL..78.2292W,1999ApJ...521L.117E,2001PhRvL..87q1102M,2002bjgr.conf...30M,2003PhRvL..90x1103R,2003PhRvD..68h3001R,2005PhRvL..95f1103A,2005MPLA...20.2351R,2006ApJ...651L...5M}. Emission from early GRB afterglows is also plausible (e.g., \cite{2007PhRvD..76l3001M}). The emission mechanism is likely similar for both long and short GRBs, the former expected to be the stronger emitter. Core-collapse supernovae with rapidly rotating cores \cite{2005MPLA...20.2351R,HEN2005PhysRevLett.93.181101}, magnetars \cite{2003ApJ...595..346Z,2005ApJ...633.1013I} and millisecond protomagnetars \cite{2009PhRvD..79j3001M} are also thought to emit high-energy neutrinos. While recent limits obtained with the IceCube detector constrain some emission models \cite{2012Natur.484..351A}, the standard fireball picture for neutrino emission remains viable \cite{PhysRevLett.108.231101,2012ApJ...752...29H}.

As their production mechanism is connected to that of gamma photons, high-energy neutrinos are probably beamed similarly to that of gamma ray emission. This decreases the rate of detectable joint sources similarly to the case of electromagnetic emission.

For a joint detection, an especially interesting class of sources are GRBs with little or no detectable electromagnetic counterpart, which are therefore only detectable via GWs and neutrinos. Such sources include so-called \emph{choked} GRBs. In the core-collapse supernova scenario, gamma-ray emission is observable only if the relativistic outflow from the central engine, that is responsible for the production of gamma-rays, breaks out of the star \cite{2008PhRvD..77f3007H,2012PhRvD..85j3004B}. The outflow, in order to advance, needs to be driven by an (active) central engine. If the breakout time of the outflow is longer than the duration of the active central engine, the outflow stalls, producing no observable gamma-ray emission \cite{2001PhRvL..87q1102M}. High-energy neutrinos are produced within the outflow and can escape through the stellar envelope, producing similar neutrino emission as for ``successful'' GRBs. Other interesting sources include low-luminosity (LL) GRBs, that can frequently go unobserved if their luminosity falls below the threshold of gamma-ray telescopes. LL GRBs, while weaker neutrino sources than their high-luminosity (HL) counterparts, are thought to be much more frequent, making their overall neutrino flux comparable or even surpass the flux from conventional HL GRBs ~\cite{2006ApJ...651L...5M,2007APh....27..386G,2007PhRvD..76h3009W}. Besides being interesting for sources not observed through their electromagnetic emission, GW-high-energy neutrino observations can provide information on the internal structure of the progenitor, as well as the properties and dynamics of the relativistic outflow (e.g., \cite{2003PhRvD..68h3001R}).

A potentially interesting subclass of GRBs for joint GW-high energy neutrino observations is low-luminosity GRBs \cite{2006Natur.442.1014S,Liang2007,2006ApJ...651L...5M,2011ApJ...739L..55B} that may be a distinct population from high-luminosity GRBs \cite{Liang2007,2007ApJ...659.1420T,2011ApJ...739L..55B}. It was suggested that more massive stars may form a black hole after collapse, and produce a high-luminosity, high-Lorentz factor outflow, while less massive stars may form a neutron star after collapse, and drive an outflow with low luminosity and Lorentz factor \cite{2006Natur.442.1018M,2007ApJ...659.1420T}. More recently, Bromberg et al. \cite{2011ApJ...739L..55B} found that estimated jet breakout times for low-luminosity GRBs are much longer than the observed durations, in marked difference with high-luminosity GRBs, suggesting that low-luminosity GRBs may be ``choked", i.e. their relativistic outflow stalls before breaking out of the stellar envelope (see also \cite{2012arXiv1206.0700P,2012ApJ...749..110B}). Low-luminosity GRBs are thought to be an order of magnitude more common than their high-luminosity counterparts, with lower electromagnetic luminosity and smaller beaming angle \cite{2006Natur.442.1014S,Liang2007}, making them a promising source type for multimessenger detections.

Joint searches for GWs and high-energy neutrinos are well-suited for multimessenger analyses, since both messengers have typically sub-threshold significances. For this purpose Baret et al. \cite{2012PhRvD..85j3004B} developed a multimessenger analysis method that combines the significance and other information from the two messengers.

\subsection{Host galaxies}
\label{subsection:hostgalaxies}

The location of potential or confirmed host galaxies of GRBs can be utilized in GW searches to enhance sensitivity as well as in event selection or localization.
Long GRBs are found to occur in star-forming regions of distant galaxies, in accordance with their association to the deaths of massive stars \cite{2006ARA&A..44..507W}. Consequently, the rate of long GRBs is correlated to the blue luminosity of galaxies \cite{2002AJ....123.1111B,2005NewA...11..103S,2006ARA&A..44..507W,2006Natur.441..463F,2009ApJ...691..182S}. Long GRBs are actually found in the very brightest regions of their host galaxies, which are significantly fainter and more irregular than typical host galaxies of core-collapse supernovae, suggesting that they are associated with extremely massive stars, and galaxies with limited chemical evolution \cite{2006Natur.441..463F}.

The distribution of short GRBs seem to follow that of old ($\sim1\,$Gyr) stellar populations within normal star forming as well as elliptical galaxies \cite{2007PhR...442..166N,2009ApJ...690..231B,2011NewAR..55....1B}, in stark contrast with the distribution of long-GRB host galaxies \cite{2007PhR...442..166N}. It is highly unlikely that short and long GRBs are drawn from the same underlying population \cite{2009ApJ...690..231B}. Nevertheless, the rate of short GRBs in a given galaxy is correlated with the optical light of their host, and to less extent with the blue luminosity of the galaxy \cite{2010ApJ...708....9F}, indicating that short-GRB progenitors have a wide age distribution of several Gyr. As neutron stars may experience an initial kick at birth \cite{1998AA...332..173P,1998ApJ...496..333F,2002ApJ...579L..63D}, binary mergers can take place far away from the star-forming region where they originate from. For kick velocities of $\mathcal{O}(100\,\mbox{km}\,\mbox{s}^{-1})$ and inspiral times of several Gyr, binaries could travel large distances before their merger \cite{2007PhR...442..166N,2010ApJ...725L..91K}, traveling far outside their host galaxies. Typical predicted distances are $\sim10-100\,$kpc \cite{2007ApJ...664.1000B}, the predicted distance distribution being well-matched by observed short-GRB distributions \cite{2010ApJ...722.1946B}.

Magnetars \cite{1995MNRAS.275..255T,2007PhR...442..166N} form another possible progenitor type of short GRBs, making up less than $\sim$one-third of their population \cite{2009ApJ...690..231B}. They are weaker and softer gamma-ray emitters than other short GRBs, making them identifiable mostly at smaller distances, within the Milky Way and the Large Magellanic Cloud.
Magnetars are thought to be created in supernova explosions \cite{2011AdSpR..47.1326H}. As they can receive an initial kick velocity during the supernova explosion similarly to radio pulsars, only a fraction of them may be near its respective supernova remnant, in accordance with observations \cite{2011AdSpR..47.1326H}. Magnetars are observed to be mostly young ($\sim$10$^4$~yr) objects \cite{2011AdSpR..47.1326H}, therefore they did not travel far from the star forming region of their respective supernovae. Consequently, extragalactic magnetar flares can be expected to occur within the star-forming regions of galaxies.

The expected distribution of GRBs can be used to enhance GW searches through identifying preferential GW source directions (e.g., \cite{2012PhRvD..85j3004B,2012arXiv1210.6362N}). A priori information on source distribution has been used in multiple GW-GRB searches \cite{070201,2012arXiv1205.1124E,2012A&A...541A.155A,2012ApJ...755....2A}.

The galaxy distribution plays a significant role in electromagnetic follow-up searches. Since most GW signal candidates have a reconstructed direction with an uncertainty of a few degrees (which may be scattered over a larger area on the sky), the sky area of the possible GW source direction is greater than the field-of-view of most astronomical instruments used for follow-up observations. To decrease the surveyed area on the sky, the directions in which electromagnetic follow-up is performed is down-selected using the directions of galaxies overlapping with GW sky area of the possible GW source direction \cite{2008CQGra..25r4034K}.

In multimessenger searches, the distribution of galaxies can be used to increase the significance of astrophysical signals as well as to reject background events whose directions do not overlap with the direction of a galaxy (e.g., \cite{2012PhRvD..85j3004B}). The significance of event candidates can be enhanced by weighing events with the a priory probability distribution of multimessenger sources (e.g., based on the luminosity and distance of galaxies).

Galaxy sky locations are especially important for GW sources that are detectable only within $\sim100\,$Mpc. On this distance scale, existing galaxy catalogs are more complete (e.g., \cite{0264-9381-28-8-085016}), and the number of galaxies is small within the sky area defined by the directional uncertainty of GW observations ($\sim$ few degrees; \cite{2012arXiv1210.6362N}). The use of galaxy direction on this scale could therefore significantly add to the sensitivity of these searches. Galaxy catalogs can still aid searches for sources at farther distances, although the sensitivity gain significantly reduces with distance \cite{2012arXiv1210.6362N}.

\section{Discussion \& Outlook}
\label{section:conclusion}

In this review we presented an overview of the major directions in which gravitational-wave (GW) astronomy has the potential of advancing our understanding of GRBs and their progenitors. We discussed the different promising directions that can lead to GW emission from GRB progenitors that is detectable with second-generation GW observatories. We also discussed some of the major scientific questions that could be addressed upon the detection of GWs from these sources.
Second generation GW observatories, commencing their operation in the next few years (e.g., \cite{LVCcommissioning}), will be sensitive enough to reach distances for which the detection of GW signals from GRB progenitors becomes possible.

The GW signal expected from compact binary coalescences, the likely progenitors of short GRBs, is fairly well understood as a function of the (often unknown) source properties. The detection of GWs can provide estimates or constraints on these source properties. We presented a detailed overview of the questions that can be addressed about short-GRB progenitors upon the detection of their GW signature.

The core-collapse of massive stars, the likely progenitor of long GRBs, has less constrained theoretical GW emission models. This is partly due to the lack of our detailed understanding of the progenitor (e.g., how cores maintain their high rotation rate prior to collapse), and partly due to the complex dynamics of the collapsing core and the resulting compact object (protoneutron star or black hole) and accretion disk, whose understanding requires further theoretical and numerical studies. Nevertheless, the core-collapse dynamics may lead to GW signals that are detectable on distance scales comparable to the lower end of the long-GRB distance scale ($\gtrsim100\,$Mpc). For example rotational instabilities in differentially rotating protoneutron-stars, as well as global instabilities in tori around black holes, both lead to rotating, non-axisymmetric systems with potentially strong GW emission.

Starquakes in highly-magnetized neutron stars, likely responsible for producing a fraction of short GRBs, result in seismic oscillations in the neutron star, which in turn may produce GWs detectable with second generation GW observatories. The detection (or non-detection) of such GWs coincident with the flaring activity of neutron stars can be informative, e.g., on the properties of matter at nuclear densities, as well as on the properties of astrophysical neutron stars.


The authors gratefully acknowledge the excellent comments, help, and suggestions of Alessandra Corsi, Jolien Creighton, Riccardo DeSalvo, Raymond Frey, Kunihito Ioka, Jonah Kanner, Kostas Kokkotas, Koutarou Kyutoku, Ilya Mandel, Brian Metzger, Kohta Murase, Ehud Nakar, Tsvi Piran, Maurice Van Putten, Luciano Rezzolla, Stephan Rosswog, Yuichiro Sekiguchi, Peter Shawhan and Masaru Shibata. The authors especially thank Richard O'Shaughnessy for his detailed comments throughout the preparation of the manuscript.
IB \& SM are grateful for the generous support of Columbia University in the City of New York and the National Science Foundation under cooperative agreement PHY-0847182. ROS \& PB are supported by NSF award PHY-0970074 and the UWM Research Growth Initiative. ROS \& PB appreciate the hospitality of the UCSB KITP, supported by the NSF with award NSF PHY11-25915.

\section*{References}
\bibliographystyle{iopart-num}

\providecommand{\newblock}{}

\end{document}